\documentclass[amsfonts,amsmath,showpacs,prb]{revtex4}

\usepackage{graphicx}
\usepackage{dcolumn}
\usepackage{bm}

\begin{document}

\title{Frustrated two-level impurities in two-dimensional antiferromagnets}

\author{A. V. Syromyatnikov}
 \email{syromyat@thd.pnpi.spb.ru}
\author{S. V. Maleyev}
\affiliation{Petersburg Nuclear Physics Institute, Gatchina, St.\ Petersburg 188300, Russia}

\date{\today}

\begin{abstract}

Dynamical properties of the impurity spin-$\frac12$ in 2D and quasi-2D Heisenberg antiferromagnets (AFs) at $T\ge0$ are discussed. The specific case of an impurity coupled {\it symmetrically} to two neighboring host spins is considered. The specific feature of this problem is that the defect is degenerate (frustrated) being located in zero molecular field. It is shown that this problem can be described by spin-boson model without tunneling term and with a more complex interaction. We demonstrate that the effect of the host system on the defect is completely described by the spectral function. It is found within the spin-wave approximation that for not too small $\omega$ the spectral function is proportional to $\omega^2/J^3$, where $J$ is the exchange constant between the host spins. The defect dynamical susceptibility is derived using Abrikosov's pseudofermion technique and diagrammatic expansion. The calculations are performed within the fourth order of the dimensionless coupling parameter $f$. It is found that transverse impurity susceptibility $\chi_\perp(\omega)$ has a Lorenz peak with the width proportional to $f^4J(T/J)^3$ which disappears at $T=0$, and a non-resonant term. The later term diverges logarithmically as $\omega,T \to 0$. The static susceptibility $\chi(0)$ has the free-spin-like contribution $1/(4T)$, and a logarithmic correction proportional to $f^2\ln(J/T)$. The influence of finite concentration of the defects $n$ on the low-temperature properties of AF is also investigated. A logarithmic correction to spin-wave velocity of the form $nf^4\ln|J/\omega|$ and an anomalous damping of spin waves proportional to $nf^4|\omega|$ are obtained. The results of the present paper can be applied to other systems with a frustrated impurity in which the spectral function is proportional to $\omega^2$.

\end{abstract}


\maketitle

\section{Introduction}

Defects (or impurities) with intrinsic degrees of freedom exist in many condensed matter systems. Isolated spin in metal is the most known example (Kondo impurity). There is a number of other types of such impurities including two-level systems in glasses, crystal-field states of the rare-earth ions, degenerate or slightly split Jahn-Teller defects, quantum dots etc. The interaction of these impurities with propagating excitations of the host system (electrons in metals, phonons, spin-waves etc.) governs the impurity dynamics and the low-temperature thermal and transport properties of the host system. 

The widely used model for investigation of two-level defect dynamics is the spin-boson model which Hamiltonian has the form: \cite{leggett,weiss} 
\begin{equation}
\label{ham1}
{\cal H}_{sbm} = {\cal H}_d + {\cal H}_0 + {\cal H}_{int},
\end{equation}
where the first, the second and the third terms describe, respectively, the isolated defect, the host system and their interaction. Then, ${\cal H}_d = -\frac12 \Delta \sigma_x + \frac12 \varepsilon \sigma_z$, where $\mbox{\boldmath $\sigma$}$ is the Pauli vector describing the defect, ${\cal H}_0$ is modeled by a set of harmonic oscillators: ${\cal H}_0 = \sum_\alpha [\frac12 m_\alpha \varepsilon_\alpha Q_\alpha^2 + P_\alpha^2/(2m_\alpha)]$ and the interaction term has the form ${\cal H}_{int} = \sigma_z\sum_\alpha C_\alpha Q_\alpha$, where $C_\alpha$ are some constants. Essentially, dynamics of the defect is determined by the spectral function characterizing the system. Commonly a power-law dependence $\omega^v$ of this function is discussed, where $v\ge0$. In the most investigated Ohmic case $v=1$. Despite its simplicity the spin-boson model has found numerous applications ranging from electron transfer to quantum information processing. \cite{leggett,weiss} Meanwhile its modifications are needed in some cases. 

In Ref.~\cite{mal2} one of us (S.V.M.) has extensively studied the problem of interaction of a defect with intrinsic degrees of freedom with 3D acoustic phonons in dielectrics. A specific approach has been proposed in which degeneracy of the impurity is assumed to be arbitrary. This approach is based on Abrikosov's pseudofermion technique \cite{abrikos} and diagrammatic expansion. In the case of the two-level defect the Hamiltonian of the model considered in Ref.~\cite{mal2} differs from Eq.~(\ref{ham1}) by the absence of ${\cal H}_d$ (the defect is assumed to be degenerate) and by another type of interaction which has the more general form 
\begin{equation}
\label{int}
{\cal H}_{int} = g \sum_\mu S^\mu\epsilon^\mu({\bf R}_0),
\end{equation}
where ${\bf S} = \frac12\mbox{\boldmath $\sigma$}$, ${\bf R}_0$ determines the position of the impurity in the crystal, $g$ is the interaction strength, index $\mu$ labels Cartesian components, and $\epsilon^\mu({\bf R}_0)$ are some operators of the host system. It was found in Ref.~\cite{mal2} that similar to the spin-boson model the effect of the host system on the defect is completely described by the spectral function given by the imaginary part of the retarded Green's function of operators $\epsilon^\mu({\bf R}_0)$:
\begin{equation}
\label{deltaf}
\Delta_{\mu\nu}(\omega) = -i\int_0^\infty dt e^{i\omega t} \langle [\epsilon^\mu({\bf R}_0,t),\epsilon^\nu({\bf R}_0,0)]\rangle,
\end{equation}
where $\langle\dots\rangle$ denotes the thermal average. In the case of 3D acoustic phonons ${\rm Im}\Delta_{\mu\nu}(\omega)$ is proportional to $\omega^3$. The approach proposed in Ref.~\cite{mal2} allows to obtain all the results in a general form independent of the particular view of the operators $\epsilon^\mu({\bf R}_0)$ and the value of $S$. The only restriction is that the spectral function is proportional to $\omega^3$. The results for various systems would differ only by some constants. Then, one can discuss the effect of the new terms in the interaction in comparison with the spin-boson model. In the case of interaction (\ref{int}) all the components of the susceptibility have $T$-independent non-resonant term and a Lorenz peak with the width $\varGamma\propto f^4(T/\Theta)^5$, where $f$ is the dimensionless coupling constant and $\Theta$ is a characteristic energy. The real part of the non-resonant term is a constant at $|\omega|\ll\Theta$ and the imaginary one is proportional to $\omega$. At the same time in the spin-boson model the transverse susceptibility has only the non-resonant term. It was also shown \cite{mal2,kok} that the scattering on the impurities leads to the anomalous 3D acoustic phonon damping proportional to $nf^4\omega^2$, where $n$ is concentration of the impurities (damping caused by scattering on static defects is proportional to $\omega^4$). Afterward the suggested approach has been successfully applied to investigation of defects in glasses \cite{mal4} and in cubic metals. \cite{mal5}

\begin{figure}
\centering
\includegraphics[scale=0.5]{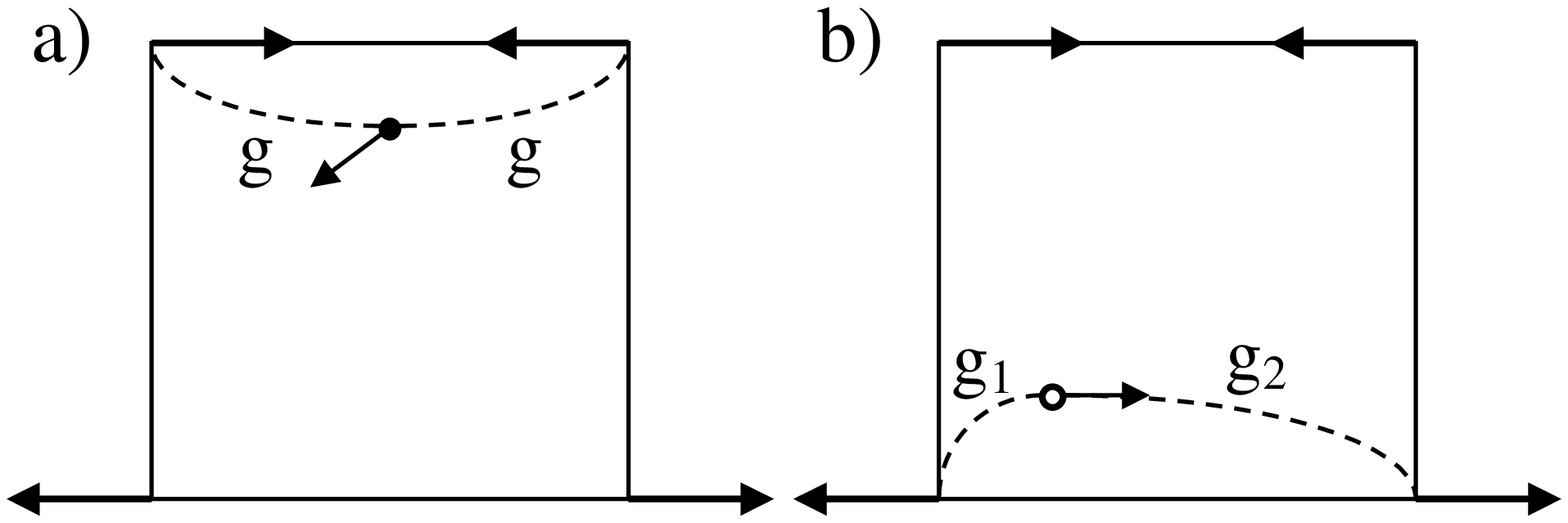}
\caption{Unit cells of 2D AF with (a) symmetrically and (b) asymmetrically coupled impurity spins are presented. Strengths of coupling with corresponding host spins $g$ and $g_1\ne g_2$ are depicted. The local Neel order is also shown. Only symmetrically coupled impurities are discussed in this paper.
\label{pic}} 
\end{figure}

In the present paper we apply the approach discussed in Ref.~\cite{mal2} with the spectral function proportional to $\omega^2$ to the problem of two-level degenerate defect. As mentioned above, the nature of the defect and the host system is not essential. The results will depend on the special form of $\epsilon^\mu({\bf R}_0)$ in Eq.~(\ref{int}) via some constants. Thus, our discussion are applicable to all systems with degenerate defect and the spectral function proportional to $\omega^2$. We demonstrate below that an example of such a system is 2D Heisenberg antiferromagnet (AF) with the impurity spin-$\frac12$ coupled {\it symmetrically} to two neighboring host spins (see Fig.~\ref{pic}). This is the particular subject of the present investigation. 

Below we show that the spectral function is proportional to $\omega^2$ if the interaction of the defect with 2D AF is determined by spin waves. It is well known that there is no long range order in Heisenberg 2D AF at $T>0$. \cite{mw} Nevertheless, as it has been shown theoretically \cite{kop,chak,tyc} and confirmed experimentally \cite{thur}, the spin waves are well defined in paramagnetic phase of 2D AF if their wavelength is much smaller than the correlation length $\xi\propto \exp({\rm const}/T)$. It is found below that the interaction is determined by spin waves and the spectral function is proportional to $\omega^2$ if $\omega\gg Ja/\xi$, where $J$ is the coupling constant between the host spins and $a$ is the lattice constant. Then, a small interaction (for definiteness interplane interaction) of the value of $\eta\ll J$ can stabilize the long range order at finite $T$. It is obtained below that the spectral function is proportional to $\omega^2$ at $\omega\gg\eta$ for the ordered quasi-2D AF. We assume that the interaction of the defect with AF has the form (\ref{int}) with $\epsilon^\mu({\bf R}_0) = s_1^\mu+s_2^\mu$:
\begin{equation}
\label{intaf}
{\cal H}_{int}^{AF} = g {\bf S} ({\bf s}_1 + {\bf s}_2),
\end{equation}
where ${\bf s}_{1,2}$ denote the host spins from different sublattices. For the following consideration the sign of $g$ is insignificant. 

It should be stressed that one must distinguish symmetrically and asymmetrically coupled impurities (see Fig.~\ref{pic}). Symmetrically coupled impurity is located in the zero molecular field. It remains degenerate and the spectral function is proportional to $\omega^2$. In the case of asymmetrically coupled impurity, where the molecular field is nonzero, there is splitting of the impurity levels and the spectral function has terms with weaker $\omega$-dependence. For instance, we demonstrate below that the spectral function for defect coupled to one host spin is proportional to a constant. In this paper we consider only the symmetric case. Our results are also valid with certain additional restrictions for slightly split nearly symmetrically coupled impurities (see below). 

Previously, different types of impurities in 2D Heisenberg AF have been extensively studied. It is believed that this problem has a relevance to the physics of some high-$T_c$ materials. In such compounds as La$_{2-x}$Sr$_x$CuO$_4$ and YBa$_2$Cu$_3$O$_{6+x}$ the mobility of the holes is very small at low level of doping before the onset of superconductivity. This finding has generated particular interest to problems of an extra spin coupled to one spin of 2D AF, \cite{hog1,hog2,sush1,sush2,sachdev,vojta,nagaosa,igar,murayama,clarke} and an extra spin coupled to sublattices symmetrically. \cite{oitmaa,kot2,clarke} It was proposed in Ref.~\cite{aharony} that the appearance of the hole could lead to a ferromagnetic interaction between corresponding two spins of the lattice. This work has stimulated studies of a missing or a ferromagnetic bond between one pair of spins in the lattice. \cite{hog2,aris} As inserting the static non-magnetic defects into the planes by means of replacing of Cu atoms with non-magnetic ones (e.g., Zn) is a common method to investigate the properties of CuO$_2$ planes of high-$T_c$ compounds, a missing spin in the lattice (vacancy) has been also discussed. \cite{hog2,chern,wan,muc,sachdev,vojta} 

The problem of an added spin in 2D AF at $T=0$ has been studied theoretically in Ref.~\cite{nagaosa}. One of the most remarkable findings of that paper is a singular logarithmic frequency behavior of the defect dynamical susceptibility. Impurity static magnetic susceptibility $\chi(\omega=0)$ for 2D AF has been evaluated in Ref.~\cite{clarke} in the case of symmetrically coupled impurity. It was demonstrated that $\chi(0)$ has a Curie-like term and a singular logarithmic correction proportional to $g^2\ln(J/T)$. 

The defect static magnetic susceptibility in 2D AF being near quantum critical point (QCP) has been discussed recently in Refs.~\cite{sachdev,vojta} using nonlinear sigma model. A classical-like behavior of the form $\chi(0)=S^2/(3T)$ with a logarithmic correction proportional to $\ln(1/T)$ was obtained. The constants before and under the logarithm were found to be universal near QCP, being the same for all types of defects and independent of the strength of the defect coupling. It is argued in Ref.~\cite{sachdev} that this behavior of static susceptibility also holds far from QCP for vacancy and for impurity spin coupled to one host spin when $T\ll |g|$. Meanwhile the constants are no longer universal far from QCP. This finding is in agreement with results of numerical simulations. \cite{hog1,hog2} They have been also confirmed by some other theoretical approaches. \cite{sush2} Both the classical-like form of $1/T$-term and the logarithmic correction are related to the nontrivial long-range dynamics in the system. \cite{sush2,sachdev}

In the present paper we study Heisenberg 2D AF at $T\ge0$ far from QCP. Our aim is to find the dynamical susceptibility of the impurity $\chi(\omega)$ at $T\ge 0$ and to discuss the influence of such impurities on the low-$T$ properties of 2D AF. These problems have not been addressed yet for symmetrically coupled defects: only the ground state properties at $T=0$ \cite{kot2,oitmaa} and the static susceptibility \cite{clarke} have been discussed. The calculations are performed within the order of $f^4$, where $f\propto g/J$ is the dimensionless coupling parameter. We show that the transverse impurity susceptibility $\chi_\perp(\omega)$ has a Lorenz peak with the width $\varGamma\propto f^4J(T/J)^3$ that disappears at $T=0$, and a non-resonant term. The imaginary part of the non-resonant term is a constant independent of $T$ at $|\omega|\gg\varGamma$ and the real part has a logarithmic divergence as $\omega,T\to0$. Similar logarithmic singularity was found in Ref.~\cite{nagaosa} at $T=0$. The longitudinal susceptibility $\chi_\|(\omega)$ has the non-resonant term which differs from that of $\chi_\perp(\omega)$ by a constant and a Lorenz peak. We demonstrate that within the order of $f^4$ the width of the peak is zero. Its calculation is out of the scope of this paper.

The static susceptibility has the free-spin-like term $S(S+1)/(3T)$ and a correction proportional to $f^2\ln (J/T)$. We point out here the sharp difference between symmetrically and asymmetrically coupled impurities that takes place in the regime $T\ll |g|$ (by asymmetrically coupled impurities we mean here either the added spin coupled to one host spin or the vacancy which is the particular case of the added spin with $g\to\infty$). The leading $1/T$-term has the free-spin-like form in the symmetric case and the classical-like form in the asymmetric one. Moreover, the logarithmic correction is proportional to $g^2$ in the symmetric case and it does not depend on $g$ in the asymmetric one. \cite{sush2,sachdev} The difference is related to the fact that the impurity spin coupled asymmetrically aligns with the local Neel order, \cite{sush2,sachdev} whereas the symmetrically coupled impurity is located in the zero molecular field.

The fact that the spectral function in 2D AF is proportional to $\omega^2$ only at $\omega\gg\{\eta \mbox{ \rm or } Ja/\xi\}$ leads to the following restriction on the range of validity of the results obtained: $\max\{\varGamma,|\omega|\}\gg\{\eta \mbox{ \rm or } Ja/\xi\}$. If the defect is slightly split (for definiteness by magnetic field $\bf H$) this condition turns into $\max\{\varGamma,|\omega|\}\gg\max\{\{\eta \mbox{ \rm or } Ja/\xi\},g\mu_BHS\}$. For nearly symmetrically coupled impurity one has: $\max\{\varGamma,|\omega|\}\gg\max\{\{\eta \mbox{ \rm or } Ja/\xi\},|g_1-g_2|\}$, where $g_{1,2}$ are values of coupling with the corresponding sublattices (see Fig.~\ref{pic}).

The results described above are valid for isotropic interaction (\ref{intaf}). We also consider interaction containing only one term: ${\cal H}_{int} = gS^x(s_1^x+s_2^x)$. In this case the $xx$-component of the impurity susceptibility is zero whereas $yy$- and $zz$- ones have only the non-resonant term. This model is identical to the spin-boson model (\ref{ham1}) without ${\cal H}_d$. The Hamiltonian can be diagonalized exactly and an exact expression for $\chi(\omega)$ can be obtained. \cite{pirc} Below we perform the corresponding calculations for the spectral function proportional to $\omega^2$ and confirm the results obtained within our approach. One of the most interesting features of the exact result is that the static susceptibility has the form $\chi(0) \propto T^{-1-\zeta}$, where $\zeta\propto f^2T/J$. Within the first order of $f^2$ one has $1/(4T)$-term and the logarithmic correction. Thus, we see that in the modified spin-boson model taking into account the higher order logarithmic corrections leads to the non-trivial power-law $T$-dependence of $\chi(0)$.

The influence of the finite concentration $n$ of the defects on the low-temperature properties of 2D AF is also considered. For not too small $\omega$ we find the logarithmic correction to the spin-wave velocity of the form $nf^4\ln|J/\omega|$ and an anomalous damping of the spin-waves proportional to $nf^4|\omega|$. Similar logarithmic correction to the velocity and damping were obtained in Ref.~\cite{chern}, where vacancies in 2D AF were studied. It is demonstrated that interaction of the spin waves with defects modifies the spectral function which acquires new terms proportional to $n$ exhibiting weaker $\omega$-dependence. These terms should be taken into account at small enough $\omega$ and the problem should be solved self-consistently. The corresponding consideration is out of the scope of this paper. Within the range of validity of our study we do not obtain a renormalization of the magnetic specific heat which is proportional to $T^2$ in 2D AF without impurities. At the same time it was obtained \cite{chern,muc} that vacancies give rise to a constant contribution to the density of states that in turn leads to a large correction to the specific heat proportional to $nT$.

The rest of the paper is organized as follows. The model, Abrikosov's pseudofermion and diagrammatic techniques employed for the calculations are discussed in Sec.~\ref{mat}. The pseudofermion Green's function, the pseudofermion vertex and the impurity dynamical susceptibility are derived in Secs.~\ref{pfgf}--\ref{suspr}. Another type of interaction of the defect with the host system are discussed in Sec.~\ref{exact}. The exactly solvable spin-boson model (without ${\cal H}_d$) which is a special case of our model is also studied in Sec.~\ref{exact} and a comparison with our results is made. Influence of the defects on low-temperature properties of 2D AF is considered in Sec.~\ref{influence}. The spin-wave spectrum and the specific heat are studied in detail in Sec.~\ref{influence}. Section~\ref{concl} contains our conclusions. A few appendices are included with details of the calculations.

\section{Model and technique}
\label{mat}
\subsection{Model}
\label{model}

Let us formulate in somewhat detail the model discussed in this paper. We consider systems which Hamiltonian can be represented in the following form:
\begin{equation}
\label{ham}
{\cal H} = \sum_{\bf k}\epsilon_{\bf k}\left(\alpha^\dagger_{\bf k}\alpha_{\bf k}+\frac12\right) + {\cal H}_{int},
\end{equation}
where the first term describes non-interacting low-energy propagating modes of the host system (e.g., phonons or magnons) and the second one is the coupling of the degenerate impurity with the system for which we have the general expression (\ref{int}). It is also supposed in this paper that the imaginary part of the function $\Delta_{\mu\mu'}(\omega)$ given by Eq.~(\ref{deltaf}) has the form
\begin{equation}
\label{delta}
{\rm Im}\Delta_{\mu\nu}(\omega) = -A \left(\frac{\omega}{\Theta}\right)^2 {\rm sgn}(\omega)\Lambda(\omega)d_{\mu\nu},
\end{equation}
where $A$ is a positive constant which dimensionality is inverse energy, $\Theta$ is the characteristic energy, $\Lambda(\omega)$ is a cut-off function which is equal to unity at $|\omega| < \Theta$ and decreases rapidly to zero outside this interval and $d_{\mu\nu}$ is a tensor. As was also mentioned above, the particular nature of the defect and the host system are not essential in our study. We will use the general expressions (\ref{int}) and (\ref{delta}) in all calculations. Therefore results for different systems would differ only by some constants. Nevertheless we discuss now the specific system, Heisenberg 2D AF, which can be described by this model.

It is well known that 2D AF at $T\ne0$ has no long range order. \cite{mw} The average $z$-component of the spin in AF is given by
\begin{equation}
\label{sz}
\langle s^z \rangle = s - \frac1N \sum_{\bf k} \frac{4sJ - \epsilon_{\bf k}}{2\epsilon_{\bf k}} -\frac{4sJ}{N} \sum_{\bf k} \frac{N(\epsilon_{\bf k})}{\epsilon_{\bf k}},
\end{equation}
where $N$ is the number of spins in the lattice, $s$ and $J$ are values of the spin and the exchange, respectively, $\epsilon_{\bf k}$ is the spin-wave energy which is equal to $\sqrt8sJk$ at small $k$ and $N(\epsilon_{\bf k}) = (e^{\epsilon_{\bf k}/T}-1)^{-1}$. The first term in Eq.~(\ref{sz}) gives the well known correction to the average spin at $T=0$ which is approximately equal to 0.2. The last term describes the spin reduction due to thermal fluctuations. At $\epsilon_{\bf k} \ll T$ we have $N(\epsilon_{\bf k})\approx T/\epsilon_{\bf k}$. Thus the last term diverges logarithmically at small $\bf k$ in 2D AF. A weak interaction of the value of $\eta\ll J$ such as anisotropy or an interplane interaction can screen this divergence and stabilize the long range order. For definiteness we consider interplane interaction. It leads the momentum to become a 3D vector. As a result an additional term appears in the spin-wave energy proportional to $\eta k_\perp$ at small $k_\perp$, where ${\bf k}_\perp$ is the component of the momentum perpendicular to the plane of the lattice. As a result the last term in Eq.~(\ref{sz}) is small and the spin waves are well defined if
\begin{equation}
\label{gencond}
\frac {T}{sJ} \ln \left( \frac{T}{s\eta} \right)\ll 1.
\end{equation}
We will assume below that this condition holds.

It is shown in Appendix~\ref{delprop} that within the spin-wave approximation the function ${\rm Im}\Delta_{\mu\nu}(\omega)$ has the form (\ref{delta}) for 2D Heisenberg AF when $|\omega|\gg \{\eta \mbox{ \rm or } Ja/\xi\}$. The particular expressions for $\Theta$, $A$, $d_{\mu\nu}$ are also established in Appendix~\ref{delprop}. Within the spin-wave approximation the only nonzero components of $d_{\mu\nu}$ are $xx$- and $yy$- ones provided that $z$-axis is directed along magnetization of the sublattices.

Abrikosov's pseudofermion representation of the impurity spin $\bf S$ is used below. The value of $S$ is assumed to be arbitrary in this approach. Nevertheless we restrict ourself in this paper by $S=1/2$. It is demonstrated below that the matrix structure of the pseudofermion Green's function and the vertex is much simpler in this case. Consideration of larger impurity spins is out of the scope of the present paper.

We point out that a new approach has been suggested recently in Refs.~\cite{major1,major2} for spin-$\frac12$ impurity problem. This approach bases on Majorana-fermion representation of the impurity spin. It was demonstrated that this representation simplifies significantly analysis of the model if one can restrict ourselves by first terms in the expansion by coupling parameter. It will be clear soon that in our case the question of possibility of such restriction requires analysis of the diagrams of the third order within this approach. Carrying out of such analysis is out of the scope of the present paper.

\subsection{Abrikosov's pseudofermion technique}

We use below Abrikosov's pseudofermion technique for the calculation of the impurity dynamical susceptibility. It was suggested in Ref.~\cite{abrikos} for Kondo effect investigation (see also Refs.~\cite{zawadovski,larsen} for discussions). The same approach has been applied for the problem of impurity in other systems by one of us (S.V.M.) in Refs.~\cite{mal2,mal4,mal5}. Let us formulate this technique briefly in the convenient for our purpose form. 

The impurity spin $\bf S$ is represented as 
\begin{equation}
\label{s}
{\bf S} = \sum_{mm'}a^\dagger_m{\bf S}_{mm'}a_{m'},
\end{equation}
where $m$ is the spin projections, $a^\dagger_m$ and $a_m$ are operator of creation and annihilation of some particles (fermions for definiteness). It is easy to verify that the spin commutation rules are satisfied in this representation. A wave function of the impurity is characterized now by the occupation numbers of $2S+1$ states: $|n_S,n_{S-1},\dots n_{-S}\rangle$. Obviously, the states with zero or more than one particles are not physical ones and we have to eliminate them carrying out the thermodynamic average. As a result the thermodynamic average of some operator $Y$ has the form:
\begin{equation}
\label{a}
\overline{Y} = \frac{{\rm Tr}^{phys} (\rho Y)}{{\rm Tr}^{phys} (\rho)},
\end{equation}
where the traces are limited to physical states and $\rho=\exp(-{\cal H}/T)$ is the statistical operator with the Hamiltonian $\cal H$ given by Eq.~(\ref{ham}). This representation is not convenient not allowing to use the standard diagrammatic technique. To overcome this obstacle an additional term in the Hamiltonian is added: \cite{abrikos}
\begin{equation}
\label{hl}
{\cal H}_\lambda=\lambda N_{pf}=\lambda \sum_m a^\dagger_m a_m, 
\end{equation}
where $N_{pf}$ is the number of pseudofermions. We show now that the average of $Y$ has the following form equivalent to (\ref{a}):
\begin{equation}
\label{a2}
\overline{Y} = \lim_{\lambda\to\infty} \frac{{\rm Tr}(\tilde\rho Y)}{{\rm Tr}(\tilde\rho N_{pf})},
\end{equation}
where $\tilde\rho = \exp\{-\tilde{\cal H}/T\}$ and $\tilde{\cal H} = {\cal H} + {\cal H}_\lambda$. We consider in this paper such $Y$ that do not contain terms without pseudofermion operators. As a result there are no contributions both to numerator and denominator in Eq.~(\ref{a2}) from states with no particles. As $\cal H$ does not change the number of particles in state, we have for the matrix elements: $\tilde\rho_{kl}=\rho_{kl}\exp(-N_l\lambda/T)$, where $N_l$ is the number of pseudofermions in states $|k\rangle$ and $|l\rangle$. Therefore, contributions from the states with more than one particles are exponentially small compared to those from the physical states which are proportional to $e^{-\lambda/T}$. The common factors $e^{-\lambda/T}$ in the numerator and the denominator of Eq.~(\ref{a2}) cancel each other. Then contributions from states with one pseudofermion survive only in the limit of $\lambda\to\infty$ and Eqs.~(\ref{a}) and (\ref{a2}) appear to be equivalent. 

Quantities in the right part of Eq.~(\ref{a2}) can be calculated using the conventional diagrammatic technique. The Hamiltonian $\tilde{\cal H}$ in the pseudofermion representation has the form: \cite{mal2}
\begin{equation}
\label{ham2}
\tilde{\cal H} 
= 
\left(\sum_{\bf k}\epsilon_{\bf k}\left(\alpha^\dagger_{\bf k}\alpha_{\bf k}+\frac12\right) 
+
\lambda \sum_m a^\dagger_m a_m 
\right)
+ 
g\sum_{m,m',\mu}a^\dagger_{m'} S_{m'm}^\mu a_m \epsilon^\mu ({\bf R}_0) = {\cal H}_0 + {\cal H}_{int}.
\end{equation}

The dynamical susceptibility of the impurity in the representation of interaction can be written using Eqs.~(\ref{a2}) and (\ref{ham2}) in the following form: \cite{mal2}
\begin{eqnarray}
\label{chiint}
\chi_P(i\omega_n) &=& 
\lim_{\lambda\to\infty}{\cal N}^{-1}\int_0^{1/T}d\tau e^{i\omega_n\tau}{\rm Tr}\left[ e^{-{\cal H}_0/T} T_\tau \left\{P(\tau)P(0){\mathfrak S}\left(\frac1T\right)\right\}\right],\\
\label{n}
{\cal N} &=& {\rm Tr} \left[ e^{-\tilde{\cal H}/T} \sum_m a^\dagger_m a_m \right],\\
\label{p}
P &=& \sum_{mm'}a^\dagger_m  P_{mm'}a_{m'},
\end{eqnarray}
where $P$ is a spin projection. In the zeroth order of the interaction ${\cal H}_{int}$ we have ${\cal N} = (2S+1)e^{-\lambda/T}$.

\subsection{Diagrammatic technique}

\begin{figure}
\centering
\includegraphics[scale=0.9]{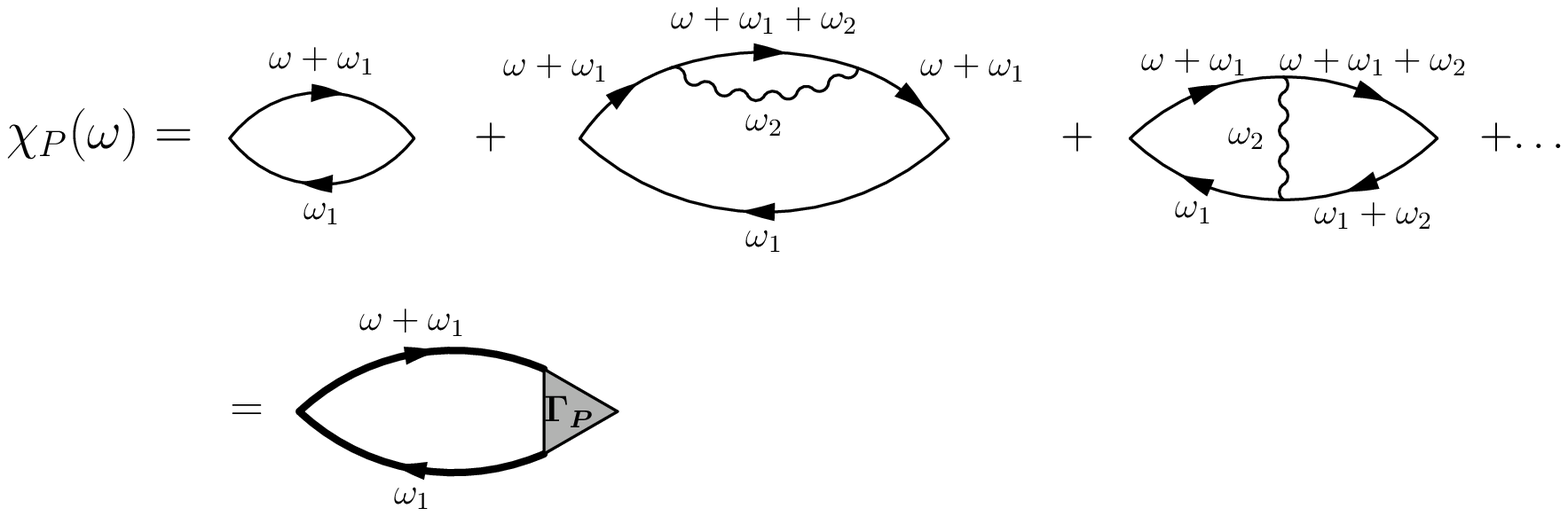}
\caption{Lower-order diagrams for the impurity dynamical susceptibility $\chi_P(\omega)$ and a graphical representation of the result of the overall series summation. Lines with arrows represent the pseudofermion Green's functions. Wavy lines denote Green's functions of operators $\epsilon^\mu ({\bf R}_0)$ of the host system (see Eq.~(\ref{int})).
\label{chifig}} 
\end{figure}

First diagrams for $\chi_P(\omega)$ and a graphical representation of the result of all diagrams summation are shown in Fig.~\ref{chifig}, where thin lines with arrows represent the bare particle Green's functions:
\begin{equation}
\label{g}
G_{mm'}^{(0)} (i\omega_n) = \frac{\delta_{mm'}}{i\omega_n-\lambda}
\end{equation}
and wavy lines denote bosons Green's functions $\Delta_{\mu\mu'}(i\omega_n)$. Notice that the diagrams with only one pseudofermion loop should be taken into account because, as is seen from Eq.~(\ref{chiint}), each loop is proportional to the small factor of $e^{-\lambda/T}$. The contribution from diagrams with one loop is finite because their factor of $e^{-\lambda/T}$ is canceled by that from $\cal N$.

For the calculation of $\chi_P(\omega)$ we use below the diagrammatic technique employed in Refs.~\cite{mal2,mal4,mal5}. Let us discuss it briefly. Firstly, we have to make an analytical continuation of diagrams for the dressed pseudofermion Green's function and for the vertex $\Gamma_P(\omega_1+\omega,\omega_1)$ from imaginary frequencies to real ones. Then, we have to express $\chi_P(\omega)$ via these quantities.

To make the first step of this program let us choose frequencies of wavy lines to be independent variables over which the summations are taken. It can be done in such a way that these frequencies are contained in arguments of $G^{(0)}$-functions with positive sign (see Fig.~\ref{chifig}). Then, each sum over a discrete frequency can be replaced by an integral over a contour enveloped the imaginary axis with an additional factor of $(2\pi i)^{-1}N(\omega)$:
\begin{equation}
T\sum_{\omega_n} \to \frac{1}{2\pi i} \oint d\omega N(\omega),
\end{equation}
where $N(\omega)=(e^{\omega/T}-1)^{-1}$ is the Plank function. \cite{mal1,mal2,mal4,mal5} The contour can be deformed so as to embrace the real axis. In evaluation of the resultant integral one should not take into account poles of $G^{(0)}$-functions because residues in these poles are proportional to $N(\lambda)\approx \exp(-\lambda/T)$. At the same time functions $\Delta_{\mu\mu'}(\omega)$ has a discontinuity on the real axis equal to $2i{\rm Im}\Delta_{\mu\mu'}(\omega)$. As a result all contour integrals can be easily transformed to those over the real axis and we lead to the following diagrammatic technique: each wavy line corresponds to $\pi^{-1}N(\omega){\rm Im}\Delta_{\mu\mu'}(\omega)$; frequencies of wavy lines should be taken so as they are contained in arguments of $G^{(0)}$-functions with positive sign; integration over all frequencies of wavy lines is taken in the interval $(-\infty,\infty)$.

One can conclude from analysis of all concrete diagrams for the vertex $\Gamma_P(i\omega_1+i\omega,i\omega_1)$ that it is an analytical function of two independent variables $i\omega_1+i\omega$ and $i\omega_1$ with cuts along the real axis. A general proof of this statement has been also given. \cite{mal1}

As a result we have for the dynamical susceptibility after the analytical continuation from the discrete frequencies to the real axis: \cite{mal2,mal4,mal5,ginz}
\begin{eqnarray}
\label{chi}
\chi_P(\omega) &=& (2\pi i {\cal N})^{-1} e^{-\lambda/T}
\int^\infty_{-\infty}
dx e^{-x/T}
 {\rm Tr} \{P\left[
G(x+\omega)\Gamma_P^{++}(x+\omega,x)G(x)
-
G^*(x)\Gamma_P^{--}(x,x-\omega)G^*(x-\omega)
\right.
\nonumber\\
&&-
\left.
G(x+\omega)\Gamma_P^{+-}(x+\omega,x)G^*(x)
+
G(x)\Gamma_P^{+-}(x,x-\omega)G^*(x-\omega)
\right]\},
\end{eqnarray}
where $G(\omega)$ is the retarded Green's function, the trace is taken over projections of the impurity spin and signs at superscript of $\Gamma_P$ denote those of imaginary parts of the corresponding arguments (e.g., $\Gamma_P^{+-}(x,y) = \Gamma_P(x+i\delta,y-i\delta)$). An energy shift by $\lambda$ has been performed during the derivation of Eq.~(\ref{chi}). As a result the Fermi function $(e^{(x+\lambda)/T} + 1)^{-1}$ has been replaced by $\exp(-(x+\lambda)/T)$ and the functions $G$ and $\Gamma_P$ no longer depend on $\lambda$. These are those functions we calculate in the next section by the diagrammatic technique. It is clear that the bare pseudofermion Green's functions in this case are $G^{(0)}_{mm'}(\omega) = \delta_{mm'}/\omega$.

\section{Dynamical susceptibility of the impurity}
\label{ds}

We derive analytical expressions for the dynamical susceptibility of the impurity in this section. Perturbation theory is used for this purpose. It can be done if the dimensionless constant
\begin{equation}
\label{f}
f^2 = \frac{g^2A}{\Theta}
\end{equation}
is small. Meanwhile we have to take into account also terms of the order of $f^4$ because the finite width of the Lorenz peak in the dynamical susceptibility arises in this order. In 2D AF we have from Eqs.~(\ref{constants}): $f^2 = (\sqrt \pi/4s)(g/J)^2$.

\subsection{Pseudofermion Green's function}
\label{pfgf}

\begin{figure}
\centering
\includegraphics[scale=0.9]{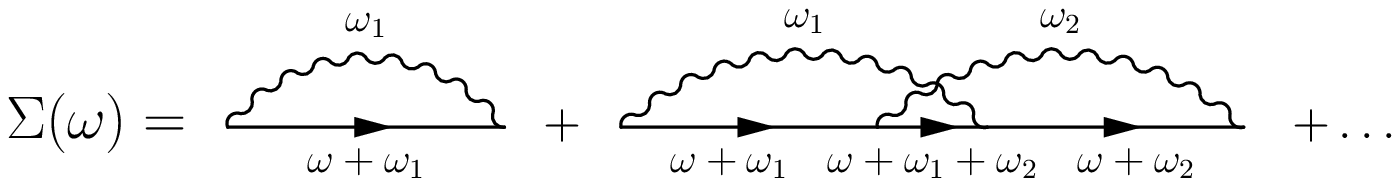}
\caption{The first- and the second-order diagrams for the self-energy $\Sigma(\omega)$ of the pseudofermion Green's function.
\label{sigmafig}} 
\end{figure}

We turn to the calculation of the pseudofermion Green's function $G_{mm'}(\omega)$. The Dyson equation for it has the following form: 
\begin{equation}
\label{dysg}
\omega G_{mm'}(\omega) = \delta_{mm'} + \sum_n\Sigma_{mn}(\omega)G_{nm'}(\omega).
\end{equation}
The first diagrams for $\Sigma_{mn}(\omega)$ are presented in Fig.~\ref{sigmafig}. Let us discuss its matrix structure first. It is determined by corresponding products of operators $S^\mu$ and tensors $d_{\mu\nu}$. For example, this combination for the second digram in Fig.~\ref{sigmafig} has the form $S^\mu S^\nu S^{\mu'} S^{\nu'} d_{\mu\mu'} d_{\nu\nu'}$. As is shown in Appendix~\ref{diag}, all such combinations are proportional to the unit matrix for the two-level impurity. It should be pointed out that there is no such simplification in the case of the impurity with the value of spin greater than 1/2. As a result the equations for the Green's function and the vertex become more complicated. The corresponding consideration of the large-spin impurities is out of the scope of the present paper.

Taking into account its matrix structure we have for the Green's function: 
\begin{equation}
\label{newg2}
G_{mm'}(\omega) = \delta_{mm'}G(\omega) = \frac{\delta_{mm'}}{\omega - \Sigma(\omega)}.
\end{equation}
The diagram of the first order shown in Fig.~\ref{sigmafig} gives the following contribution to $\Sigma(\omega)$:
\begin{eqnarray}
\label{2sig11}
\Sigma^{(1)}(\omega) &=& R_\Sigma \frac{f^2}{\pi\Theta}\int_{-\infty}^\infty dx \frac{x|x|}{x+\omega+i\delta}N(x)\Lambda(x),\\
R_\Sigma &=& S^\mu S^\nu d_{\mu\nu}
\end{eqnarray}
where the constant $f^2$ is given by Eq.~(\ref{f}) and $R_\Sigma=1/2$ for 2D AF. It is convenient to extract from this expression a term proportional to $\omega$ as follows:
\begin{equation}
\label{2sig12}
\Sigma^{(1)}(\omega) = -\omega R_\Sigma \frac{f^2}{\pi\Theta}\int_{-\infty}^\infty dx \frac{|x|}{x+\omega+i\delta}N(x)\Lambda(x) 
-
R_\Sigma \frac{f^2}{\pi\Theta}\int_{0}^\infty dx x\Lambda(x),
\end{equation}
where we have used that $\Lambda(\omega)$ is an even function and $N(-x)=-1-N(x)$. Notice that the second term in Eq.~(\ref{2sig12}) is the $T$-independent constant. Then it can be included in the renormalization of $\lambda$ and omitted. The first term in Eq.~(\ref{2sig12}) is proportional to $f^2\omega T \ln(T/\omega)$ at small $\omega$. It would seem that a great renormalization of the Green's function takes place if $f^2T \ln|T/\omega| \agt \Theta$. Meanwhile we show now that the logarithmic singularities at real $\omega$ are screened by a finite damping which is of the order of $f^4$. Let us represent the Green's function in the form
\begin{equation}
\label{2gf}
G(\omega) = \frac{1-Z(\omega)}{\omega + i \gamma(\omega)},
\end{equation}
where $Z(\omega)$ and $\gamma(\omega)$ are some functions, $\gamma(\omega)$ is a real one, and a constant term in the denominator has been attributed to the renormalization of $\lambda$ and discarded. Evaluating contribution from the first diagram in Fig.~\ref{sigmafig} using Eq.~(\ref{2gf}) we have in the first order a correction to the constant,
\begin{equation}
\label{2z}
Z^{(1)}(\omega) = R_\Sigma \frac{f^2}{\pi\Theta}\int_{-\infty}^\infty dx \frac{|x|}{x+\omega+i\gamma(x+\omega)}N(x)\Lambda(x)
\end{equation}
and $\gamma^{(1)}(\omega)=0$. The logarithmic divergence in expression (\ref{2z}) at real $\omega$ is screened by the term $i\gamma(\omega+x)$ in the denominator. It is shown below that $\gamma(\omega)$ is proportional to $f^4T^3$ at $|\omega|\ll T$. Contributions to $Z(\omega)$ from the higher order diagrams are also small by the same reason and we can restrict ourself by the first correction to it. 

To obtain $\gamma(\omega)$ one has to take into account $f^4$-terms from the first and the second diagrams shown in Fig.~\ref{sigmafig}. Along with the small corrections to the constant and to $Z(\omega)$ we have:
\begin{eqnarray}
\label{2gamm}
\gamma(\omega) &=& - R_\gamma \frac{f^4}{\pi\Theta^2}
\int_{-\infty}^\infty dx |x(x+\omega)|N(x)N(-x-\omega)\Lambda(x)\Lambda(x+\omega),\\
\label{2rg}
R_\gamma &=& S^\mu S^\nu [S^{\nu'}, S^{\mu'}] d_{\mu\mu'}d_{\nu\nu'}.
\end{eqnarray}
It is important to note that $\gamma(\omega)$ is the constant at $|\omega| \ll T$:
\begin{equation}
\label{2gamm0}
\gamma(\omega) \approx \varGamma_0 = 2\pi R_\gamma \left(\frac{f^2}{\pi\Theta}\right)^2 \int_0^\infty dxx^2N(x)[1+N(x)]\Lambda^2(x)
=
R_\gamma \frac{2\pi f^4}{3} \Theta\left(\frac T\Theta\right)^3.
\end{equation}
Notice that from the physical reason $R_\gamma$ and $\varGamma_0$ should be positive. For instance, $R_\gamma=1/4$ for 2D AF.

It is significant to note that ${\rm Im}Z(\omega)$ and $\gamma(\omega)$ have the following property at $|\omega|\gg \varGamma_0$ which will be useful in the following:
\begin{equation}
\label{prop}
\begin{array}{l}
{\rm Im}Z(-\omega) = -e^{-\omega/T} {\rm Im}Z(\omega),\\
\gamma(-\omega) = e^{-\omega/T} \gamma(\omega).
\end{array}
\end{equation}
In fact these functions are exponentially small at negative $\omega$ if $|\omega| \gg T$.

\subsection{Pseudofermion vertex}
\label{vertex}

Let us turn to the consideration of the pseudofermion vertex $\Gamma_P(x+\omega,x)$. First diagrams for this quantity are presented in Fig.~\ref{gammaf}. It is shown in Appendix~\ref{diag} that $\Gamma_{Pmm'}(x+\omega,x)$ is proportional to $P_{mm'}$. Thus, it is convenient to introduce a new quantity:
\begin{equation}
\Gamma(x+\omega,x) = \frac{\overline{P\Gamma_P(x+\omega,x)}}{\overline{P^2}},
\end{equation}
where we use the following notification: $\overline{Y} = {\rm Tr} (Y)$. As is seen from Eq.~(\ref{chi}), we need four different branches of $\Gamma(x+\omega,x)$. It is clear that $\Gamma^{++} = (\Gamma^{--})^*$ and within the first order of $f^2$ one has:
\begin{eqnarray}
\label{2g++}
\Gamma^{++}(x+\omega,x) &=& 1 + R_1\frac{f^2}{\pi\Theta}\int_{-\infty}^\infty dy y|y|N(y)\Lambda(y) G(x+y+\omega)G(x+y),\\
R_1 &=& \frac{\overline{P S^\mu PS^\nu}d_{\mu\nu}}{\overline{P^2}}.
\end{eqnarray}
It is seen from Eq.~(\ref{2g++}) that the poles of $G$-functions in the integrand are on the one hand from the real axis. Hence, the second term in Eq.~(\ref{2g++}) is much smaller than unity and we can restrict ourself by this precision. 

The situation is different in the case of $\Gamma^{+-} = (\Gamma^{-+})^*$. The first correction to it is given by Eq.~(\ref{2g++}) with $G^*(x+y)$ put instead of $G(x+y)$. Therefore, poles of the Green's functions are on the opposite sides of the real axis. As a result at $\omega=0$ the integral diverges at finite $x$ as $\varGamma_0 \to 0$ and one has to sum all series to determine $\Gamma^{+-}$. We write now an equation for $\Gamma^{+-}$ in which the most singular diagrams in each order of $f^2$ are taken into account. As a result of analysis of the diagrams up to the fourth order of $f^2$ we have obtained that in the most singular diagrams each wavy line connects points from different sides of the vertex (like in the second, the third and the fourth diagrams in Fig.~\ref{gammaf} and not like in the last one) and crosses no more than one another wavy line. Thus, the most singular diagrams are taken into account in the following equation:
\begin{eqnarray}
\label{2g+-}
\Gamma^{+-}(x+\omega,x) &=& 
1 + 
R_1 \frac{f^2}{\pi\Theta}\int_{-\infty}^\infty dy y|y|N(y)\Lambda(y) \Gamma^{+-}(x+y+\omega,x+y) G(x+y+\omega)G^*(x+y)
\nonumber\\
&&+
R_2\left(\frac{f^2}{\pi\Theta}\right)^2
\int_{-\infty}^\infty dy_1dy_2 y_1y_2|y_1y_2|N(y_1)N(y_2)\Lambda(y_1)\Lambda(y_2) 
\Gamma^{+-}(x+y_1+y_2+\omega,x+y_1+y_2) 
\nonumber\\
&&{}\times G(x+y_1+\omega) G(x+y_1+y_2+\omega) G^*(x+y_1+y_2) G^*(x+y_2),\\
R_2 &=& \frac{\overline{P S^\mu S^\nu P S^{\mu'} S^{\nu'}}d_{\mu\mu'}d_{\nu\nu'}}{\overline{P^2}}.
\end{eqnarray}
The second and the third terms in Eq.~(\ref{2g+-}) take into account diagrams with a rung and with crossing of two neighboring rungs, respectively. Let us try to solve this equation by iterations. It is easy to verify that at $|\omega|\sim \varGamma_0$ and $\varGamma_0\to0$ the divergence in the second term occurs in the second iteration only. Moreover it is of the same order as the divergence of the third term in the first iteration. As a result the equation (\ref{2g+-}) can be rewritten as follows:
\begin{eqnarray}
\label{2g+-2}
\Gamma^{+-}(x+\omega,x) &=& 
1 + 
R_1\frac{f^2}{\pi\Theta}\int_{-\infty}^\infty dy y|y|N(y)\Lambda(y) G(x+y+\omega)G^*(x+y)
\nonumber\\
&&+
\left(\frac{f^2}{\pi\Theta}\right)^2
\int_{-\infty}^\infty dy_1dy_2 y_1y_2|y_1y_2|N(y_1)N(y_2)\Lambda(y_1)\Lambda(y_2) 
\Gamma^{+-}(x+y_1+y_2+\omega,x+y_1+y_2) \nonumber\\
&&{}\times [R_1^2 G^*(x+y_1) + R_2 G^*(x+y_2)] 
G(x+y_1+\omega) G(x+y_1+y_2+\omega) G^*(x+y_1+y_2).
\end{eqnarray}
It can be solved if one notes that at $|\omega|\ll T,\Theta$ in the third term the area of integration near poles of two last Green's functions is essential. As the rest factors of the integrand changes slightly at such $y_2$ one can set in them $y_2=-x-y_1$ and neglect their dependence on $\omega$. As a result we obtain the following equation:
\begin{eqnarray}
\Gamma^{+-}(x+\omega,x) &=& 1 
+ 
R_1\frac{f^2}{\pi\Theta}\int_{-\infty}^\infty dy y|y|N(y)\Lambda(y) G(x+y+\omega)G^*(x+y)
+
K(x) \Gamma^{+-}(\omega,0) \frac{2\pi i}{\omega + 2i\varGamma_0},\\
K(x) &=& \left( \frac{f^2}{\pi\Theta}\right)^2 \int_{-\infty}^\infty dy \frac{|y (x+y )|(R_2(x+y ) - R_1^2y )}{x+y +i\gamma(x+y )}N(y )N(-x-y )\Lambda(y )\Lambda(x+y ).
\end{eqnarray}
It can be easily solved with the result:
\begin{eqnarray}
\label{gam+-tls}
\Gamma^{+-}(x+\omega,x) &=& 1 
+ 
R_1\frac{f^2}{\pi\Theta}\int_{-\infty}^\infty dy y|y|N(y)\Lambda(y) G(x+y+\omega)G^*(x+y)
+
K(x) Z_\Gamma \frac{2\pi i}{\omega + 2i\varGamma},\\
\label{vargamma}
\varGamma &=& \varGamma_0 -\pi K(0) = 2\pi R_\varGamma \left( \frac{f^2}{\pi\Theta}\right)^2 \int_0^\infty dy y^2N(y)[1+N(y)]\Lambda^2(y),\\
Z_\Gamma &=& 1 + R_1\frac{f^2}{\pi\Theta}\int_{-\infty}^\infty dyy|y|N(y)\Lambda(y)|G(y)|^2,\\
\label{rvar}
R_\varGamma &=& R_\gamma - R_1^2 + R_2 = \frac{\overline{PS^\mu S^\nu [[S^{\nu'},S^{\mu'}],P]}d_{\nu\nu'}d_{\mu\mu'}}{\overline{P^2}}.
\end{eqnarray}
In 2D AF $R_\varGamma = 1/2$ and 0 for $P = S^x,S^y$ and $P = S^z$, respectively. Evidently, the temperature dependence of $\varGamma$ for $P = S^x,S^y$ is the same as that of $\varGamma_0$ given by Eq.~(\ref{2gamm0}). Note, the third term in Eq.~(\ref{gam+-tls}) is much smaller than the second one when $|\omega|\gg \varGamma_0/f^2$.

\subsection{Properties of the impurity susceptibility}
\label{suspr}

We can derive now the impurity susceptibility using the general expression (\ref{chi}), Eqs.~(\ref{2gf}), (\ref{2z}), (\ref{2gamm}) and (\ref{2gamm0}) for the Green's function and Eqs.~(\ref{2g++}) and (\ref{gam+-tls}) for the branches of the vertex. As a result of tedious but simple calculations presented in Appendix~\ref{gameval} we have for the dynamical susceptibility of the impurity up to terms of the order of $f^2$:
\begin{eqnarray}
\label{chi2}
\chi_P(\omega) &=& \frac{\overline{P^2}}{2}
\left\{
\frac{2i\varGamma}{T(\omega+2i\varGamma)}
\left(
1 + (R_1 - 2R_\Sigma)\frac{f^2}{\pi\Theta} \int_{-\infty}^\infty dx \frac{|x|N(x)\Lambda(x)}{x + 2i\varGamma_0}
\right)
+
\frac{2i\varGamma_0}{T(\omega+2i\varGamma_0)} R_1 \frac{f^2}{\pi\Theta} \int_{-\infty}^\infty dx \frac{|x|N(x)\Lambda(x)}{x + 2i\varGamma_0}
\right.\nonumber\\
&&{}+
\left.
2R_\chi\frac{f^2}{\pi\Theta} \int_{-\infty}^\infty dx \frac{{\rm sgn}(x)\Lambda(x)}{x + \omega + 2i\varGamma_0}
\right\},\\
\label{rchi}
R_\chi &=& R_\Sigma - R_1 = \frac{\overline{PS^\mu[S^\nu, P]}d_{\mu\nu}}{\overline{P^2}}.
\end{eqnarray}
The first term in Eq.~(\ref{chi2}) is the Lorenz peak with the width $\varGamma$. The second one is a Lorenz peak with the width $\varGamma_0$ and small, proportional to $f^2$, amplitude. The last term is the non-resonant part of the susceptibility. The imaginary part of the non-resonant term in Eq.~(\ref{chi2}) at $|\omega|\gg\varGamma_0$ is proportional to ${\rm sgn}(\omega)$ and the real one contains the logarithmic singularity of the form $\ln (\omega^2 + \varGamma_0^2)$. At $T=0$ and $\omega\ne0$ the nonresonant contribution survives only and the susceptibility has the logarithmic singularity. Such a singularity has been obtained for the two-level impurity at $T=0$ in Ref.~\cite{nagaosa}. The first and the second terms in Eq.~(\ref{chi2}) are calculated assuming that $|\omega|\ll T$. At $|\omega|\gg T$ these terms are of the order of $f^4$ and their taking into account exceeds the range of accuracy. 

It should be noted once more that the particular nature of the defect and the host system is not essential in the above consideration. We use the general expression (\ref{int}) for the interaction and assume that the function ${\rm Im}\Delta_{\mu\nu}(\omega)$ has the form (\ref{delta}). Apart from $f^2$ only coefficients $R$ in the resultant expression (\ref{chi2}) depend on the nature of the defect and the host system. We demonstrate now that systems with different symmetries of the interaction show different behavior of the impurity. In the case of isotropic interaction all the components of tensor $d$ in Eq.~(\ref{delta}) are nonzero and all components of the impurity susceptibility have the same structure: the Lorenz peaks and the nonresonant term. When one of the component of $d$ is zero, say $zz$- one like in 2D AF, the behavior of transverse components $\chi_x(\omega)$ and $\chi_y(\omega)$ differs from that of longitudinal one $\chi_z(\omega)$. The transverse components have the Lorenz peak $\varGamma\sim f^4$ and the nonresonant term. The Lorenz peak with the width $\varGamma_0$ disappears because $R_1=0$. The transverse component contains the nonresonant term but $\varGamma=0$ (see Eqs.~(\ref{vargamma}) and (\ref{rvar})) and our precision is insufficient to determine the resonance terms in $\chi_z(\omega)$. The corresponding calculations of $\varGamma_0$ and $\varGamma$ with higher precision is out of the scope of the present paper. If only $d_{xx}$ is nonzero, then $\chi_x(\omega)\equiv 0$ whereas $\chi_y(\omega)$ and $\chi_z(\omega)$ have only the nonresonant term. This particular situation is considered in detail in the next section.

For static susceptibility $\chi_P(0)$ we have from Eq.~(\ref{chi2}):
\begin{eqnarray}
\label{unif}
\chi_P(0) &=& \frac{\overline{P^2}}{2T}\biggl(1 + W(T)\biggr),\\
\label{w}
W(T) &=&  2R_\chi\frac{f^2}{\pi\Theta} \int_{-\infty}^\infty dx \frac{{\rm sgn}(x)\Lambda(x)}{x + 2i\varGamma_0}
\left( T-xN(x) \right),
\end{eqnarray}
where the first and the second terms in $W(T)$ stem from the non-resonant and the resonant parts in Eq.~(\ref{chi2}), respectively. It is easy to verify that $W(T)$ is a real value up to the order of $f^2$ (remember, the susceptibility has been calculated up to terms of this order). Integrations in Eq.~(\ref{w}) can be simply carried out and we have at $T\ll \Theta$:
\begin{equation}
\label{w2}
\chi_P(0) = \frac{\overline{P^2}}{2T} \left(1 - 2R_\chi\frac{f^2}{\pi} \right) 
+
\overline{P^2} 2 R_\chi\frac{f^2}{\pi\Theta}\ln\left(\frac \Theta T\right).
\end{equation}
Thus the uniform susceptibility has the free-spin-like term $\overline{P^2}(2T)^{-1}$ which amplitude is slightly reduced by the interaction and the correction proportional to $f^2 \ln (\Theta/T)$. Expression (\ref{w2}) is in accordance with that of Ref.~\cite{clarke}. Similar result for the static susceptibility, $1/T$-term and a logarithmic singular correction to it, has been obtained in 2D AF near QCP. \cite{sachdev} It was found that the static susceptibility exhibits the classical-like Curie behavior of the form $S^2/(3T)$ and the coefficients before and under the logarithm are universal values independent of the particular type of the impurity and the strength of its coupling to the host system. Remarkably, this behavior remains also far from QCP for asymmetrically coupled impurities (vacancy and added spin) at $T\ll g$. \cite{hog1,hog2,sachdev,sush2} but the constant under the logarithm becomes non-universal. These findings are related to the nontrivial long-range dynamics of the 2D AF. Then we point out the significant difference between dynamical properties of symmetrically and asymmetrically coupled impurities in the regime $T\ll g$: the leading $1/T$-terms have the free-spin-like behavior and the classical-like one, respectively. Moreover, the logarithmic corrections is proportional to $g^2$ in the symmetric case and it does not depend on $g$ in asymmetric one. \cite{sush2,sachdev} As was also pointed out in Introduction, the difference can be explained by the fact that the impurity spin coupled asymmetrically aligns with the local Neel order \cite{sush2,sachdev} at $T\ll g$ whereas the symmetrically coupled impurity is located in the zero molecular field.

It is convenient from this point on to neglect in Eq.~(\ref{chi2}) the small corrections and use the following simple expression for the transverse susceptibility $\chi_\perp(\omega)=\chi_x(\omega)=\chi_y(\omega)$:
\begin{eqnarray}
\label{chi3}
\chi_\perp(\omega) &=& \overline{P^2}\frac{i\varGamma}{T(\omega+2i\varGamma)}
+
\overline{P^2}R_\chi\frac{f^2}{\pi\Theta} \int_{-\infty}^\infty dx \frac{{\rm sgn}(x)\Lambda(x)}{x + \omega + i\varGamma_0},\\
\label{imchi}
{\rm Im}\chi_\perp(\omega) &=& \overline{P^2}\frac{\omega\varGamma}{T(\omega^2 + 4\varGamma^2)}
+
\overline{P^2}R_\chi\frac{f^2}{\Theta} \Lambda(\omega) {\rm sgn}(\omega).
\end{eqnarray}
The imaginary part of the non-resonant term is calculated at $|\omega|\gg\varGamma_0$ in Eq.~(\ref{imchi}). We see that it does not depend on the temperature at such $\omega$. 
It is seen from Eq.~(\ref{chi3}) that the non-resonant term gives the main contribution to the susceptibility when
\begin{equation}
\label{omlim}
|\omega| \gg \omega_0 = f^2\Theta
\left( \frac T\Theta \right)^2.
\end{equation}
As the sign of ${\rm Im}\chi_P(\omega)$ should coincide with that of $\omega$, the constant $R_\chi$ given by Eq.~(\ref{rchi}) should be positive. For example, we have for 2D AF: $R_\chi=1/2$.

It should be stressed that there is a restriction on the range of validity of the resultant expressions (\ref{chi2}) and (\ref{chi3}) for $\chi_P(\omega)$ in the case of 2D AF. It is the consequence of the fact that the function ${\rm Im}\Delta(\omega)$ has the form (\ref{delta}) if $\omega \gg \{\eta \mbox{ \rm or } Ja/\xi\}$ only. It is easy to see that in all calculations performed above one can use the function of the form (\ref{delta}) if the following condition on $\omega$ and $\varGamma_0$ holds: $\max\{\varGamma_0,|\omega|\}\gg\{\eta \mbox{ \rm or } Ja/\xi\}$.

\subsection{Comparison with the exactly solvable model}
\label{exact}

Let us consider the special case when the interaction contains only one term: ${\cal H}_{int}=gS^x\epsilon^x({\bf R}_0)$. It is seen from Eqs.~(\ref{2rg}) and (\ref{rvar}) that $\varGamma=\varGamma_0=0$ and the resonant terms are zero at $\omega\ne0$ in Eq.~(\ref{chi2}). One can conclude from Eqs.~(\ref{rchi}) and (\ref{chi3}) that $\chi_x(\omega)=0$ and
\begin{equation}
\label{chi3ex}
\chi_y(\omega) = \chi_z(\omega) = \chi^{(\perp)}(\omega) = \frac{f^2}{2\pi\Theta}\ln\left|\frac\Theta\omega\right| + i\frac{f^2}{4\Theta}\Lambda(\omega){\rm sgn}(\omega).
\end{equation}
As it was pointed out above, our model in this case is equivalent to the spin-boson model (\ref{ham1}) without ${\cal H}_d$. The corresponding Hamiltonian can be diagonalized exactly and an exact expression for the impurity susceptibility can be derived. We perform in this subsection the corresponding calculations of $\chi^{(\perp)}(\omega)$ and confirm our results obtained above. 

The detailed discussion of the exactly solvable spin-boson model is also necessary by the following reason. It was important for our consideration that $\gamma(\omega)$ in expression (\ref{2gf}) for the Green's function is nonzero, i.e., the coefficient $R_\gamma$ given by Eq.~(\ref{2rg}) is finite. As is demonstrated in Sec.~\ref{pfgf}, the imaginary term $i\gamma(\omega)$ in the denominator of the Green's function $G(\omega)$ screens the logarithmic singularity of the self-energy part allowing us to restrict ourself by the second order of $f^2$. In the opposite case, when imaginary part in the denominator of $G(\omega)$ is zero, one has to sum all the series to calculate the Green's function in the region of $\omega$ determined by the condition
\begin{equation}
\label{limass}
f^2\frac T\Theta \ln \left| \frac{T}{\omega} \right| \agt 1.
\end{equation}
It should be stressed that Eq.~(\ref{chi3ex}) for $\chi^{(\perp)}(\omega)$ is valid if the condition (\ref{limass}) on the frequency $\omega$ is not fulfilled. If the condition (\ref{limass}) is fulfilled the results of the exact diagonalization should be discussed.

We represent Hamiltonian (\ref{ham1}) of the spin-boson model describing degenerate defect in the following form:
\begin{equation}
\label{hamex}
{\cal H} = \frac12 \sum_{\bf k} (\epsilon_{\bf k}^2 Q_{\bf k} Q_{-\bf k} + P_{\bf k} P_{-\bf k})
+ gS^x\sum_{\bf k} F_{\bf k} Q_{\bf k},
\end{equation}
where the symbols $\epsilon_{\bf k}$, $Q_{\bf k}$ and $P_{\bf k}$ stand for the frequency, normal coordinate and momentum of the system propagating modes (bosons) with momentum $\bf k$, where $ [Q_{\bf k},  P_{-\bf k'}] = i\delta_{{\bf kk}'}$, $Q_{\bf k} = \alpha_{\bf k} + \alpha^\dagger_{-\bf k}$, $P_{\bf k} = -i\epsilon_{\bf k} (\alpha_{\bf k} - \alpha^\dagger_{-\bf k})$. The last term in Eq.~(\ref{hamex}) describes coupling of the impurity with the system, where $F_{\bf k}$ is a coupling parameter. In this problem definition the spectral function given by Eq.~(\ref{deltaf}) has the form 
\begin{equation}
\label{exdel}
{\rm Im}\Delta(\omega) = -\frac\pi2 \sum_{\bf k}\frac{|F_{\bf k}|^2}{\epsilon_{\bf k}}[ \delta(\omega - \epsilon_{\bf k}) - \delta(\omega + \epsilon_{\bf k}) ]. 
\end{equation}
We treat the elementary excitations within Debye approximation and assume that the spectrum is linear in $k$: $\epsilon_{\bf k} = ck$. We make also one more assumption which do not effect the results: $\bf k$ is a two-dimensional vector. In this case the spectral function is proportional to $\omega^2$ when the coupling parameter is linear in $k$:
\begin{equation}
\label{fk}
F_{\bf k} = \frac{1}{\Theta \sqrt V }\epsilon_{\bf k},
\end{equation}
where $V$ is the volume of the crystal and $\Theta$ is the cut-off frequency. In this case the spectral function (\ref{exdel}) has the form (\ref{delta}) with $A = (4c^2)^{-1}$. 

The Hamiltonian (\ref{hamex}) can be diagonalized exactly. It is convenient for this purpose to apply the following canonical transformation: \cite{pirc} 
\begin{equation}
\label{hamex2}
e^{-R}{\cal H} e^R = \frac12 \sum_{\bf k} [\epsilon_{\bf k}^2 Q_{\bf k} Q_{-\bf k} + P_{\bf k} P_{-\bf k}] 
- \frac {g^2}{8} \sum_{\bf k} \frac{|F_{\bf k}|^2}{\epsilon_{\bf k}^2}, 
\qquad 
R = i g \sum_{\bf k}\frac{F_{\bf k}}{\epsilon_{\bf k}^2} P_{\bf k} S^x.
\end{equation}
It can be shown that the correlation function can be brought to the form: \cite{pirc}
\begin{eqnarray}
\label{corr}
&&\langle S^y(t) S^y(0) \rangle = \frac14 e^{I(t)},\\
\label{i}
&&I(t) = g^2 \sum_{\bf k} \frac{|F_{\bf k}|^2}{2\epsilon_{\bf k}^3}\{ N(\epsilon_{\bf k})[e^{i\epsilon_{\bf k}t} - 1] + [N(\epsilon_{\bf k}) +1] [e^{-i\epsilon_{\bf k}t} - 1] \}.
\end{eqnarray}
As a result we have for the transverse dynamical susceptibility using the representation $I(t) = X(t) - iU(t)$:
\begin{equation}
\label{chiex}
\chi^{(\perp)}(\omega) = \frac12 \int_0^\infty dt e^{i\omega t} e^{X(t)} \sin[U(t)].
\end{equation}
One obtains for $X(t)$ and $U(t)$ from Eq.~(\ref{i}):
\begin{eqnarray}
\label{x}
X(t) &=& \frac{f^2}{\pi\Theta} \int_0^\Theta dx [\cos(xt) - 1] [1+2N(x)],\\
\label{u}
U(t) &=& \frac{f^2}{\pi\Theta} \int_0^\Theta dx \sin(xt) = \frac{f^2}{\pi} \frac{1 - \cos (\Theta t)}{\Theta t},
\end{eqnarray}
where $f^2$ is given by Eq.~(\ref{f}). It is seen from Eq.~(\ref{u}) that $U(t)$ is a bounded function and we have in Eq.~(\ref{chiex}): $\sin[U(t)]\approx U(t)$. The function $X(t)$ along with negligibly small terms of the order of $f^2$ has another one which is large at $tT\gg \exp\{\Theta/(f^2T)\}$ and for which we have with the logarithmic precision: $X(t)\approx -2f^2T(\pi\Theta)^{-1}\ln(tT)$. 
\footnote{It is seen from Eq.~(\ref{corr}) that the correlation function tends to zero as $t\to\infty$. This signifies that the system is ergodic. \cite{pirc} It means, in particular, that the isolated static susceptibility (\ref{chiint}) coincides with isothermal static one $\chi^T = (\partial M/\partial H)_T$, where $M$ is the magnetization of the sample. \cite{pirc}}
As a result we obtain:
\begin{eqnarray}
\label{chiex2}
\chi^{(\perp)}(\omega) = \frac{f^2}{2\pi} \left( \frac \Theta T \right)^\zeta 
\int_0^\infty dt e^{i\omega t}\frac{1-\cos(t\Theta)}{(t\Theta)^{1+\zeta}}
&=&
\frac{1}{4T} \left( \frac \Theta T \right)^\zeta
\left[
\frac12 \left|1 + \frac \omega \Theta\right|^\zeta
+
\frac12 \left|1 - \frac \omega \Theta\right|^\zeta
-
\left|\frac \omega \Theta\right|^\zeta
\right] \nonumber\\
&&-
i\frac{f^2}{4\Theta}\left( \frac \Theta T \right)^\zeta
\left[
\frac12 \left|1 + \frac \omega \Theta\right|^\zeta
-
\frac12 \left|1 - \frac \omega \Theta\right|^\zeta{\rm sgn}(\Theta-\omega)
-
\left|\frac \omega \Theta\right|^\zeta\right] {\rm sgn}(\omega), \nonumber\\
\end{eqnarray}
where $\zeta=2f^2T(\pi\Theta)^{-1}$. We recover Eq.~(\ref{chi3ex}) from Eq.~(\ref{chiex2}) if the condition (\ref{limass}) is not fulfilled and $|\omega| \ll \Theta$ or $|\omega| \gg \Theta$. 
At $\omega\ll\Theta$ Eq.~(\ref{chiex2}) can be represented in the simple form:
\begin{equation}
\label{chiex3}
\chi^{(\perp)}(\omega) = 
\frac{1}{4T} \left( \frac \Theta T \right)^\zeta
\left[
1 -
\left( \frac{\omega}{i\Theta} \right)^\zeta
\right]
\end{equation}
At small enough $\omega$, when (\ref{limass}) holds, equation (\ref{chi3ex}) is incorrect and we see from Eq.~(\ref{chiex3}) that the susceptibility shows the nontrivial $\omega$- and $T$- dependences. In particular, the static susceptibility is proportional to $T^{-1-\zeta}$. In the order of $f^2$ the static susceptibility has the same structure as that obtained above (see Eq.~(\ref{w2})): it has the conventional term $1/(4T)$ and the logarithmic correction to this term. Thus, we see that in the modified spin-boson model taking into account higher order logarithmic corrections results in the non-trivial power-law $T$-dependence of $\chi(0)$.

\section{Influence of the defects on the host system}
\label{influence}

We discuss in this section the influence of the defects with finite concentration $n$ on the low-temperature properties of 2D AF. The spin-wave spectrum and the specific heat of AF are considered below in detail. It will be assumed that $n\ll1$ in order to neglect interaction between impurities.

\subsection{Spin-wave spectrum}

The spin-wave spectrum is determined by the poles of the spin Green's functions. The Green's functions of 2D AF with impurities are investigated in Appendix~\ref{gfwi}. It is demonstrated there that their denominator has the form
\begin{equation}
\label{sgfr}
{\cal D}(\omega,{\bf k}) = \omega^2 - \epsilon_{\bf k}^2 + 4s^2g^2n \chi_\perp(\omega) 
\bigl[ 
J_0 - J_{\bf k}\cos({\bf kR}_{12})
\bigr],
\end{equation}
where ${\bf R}_{12}$ is the vector connected host spins coupled to impurity (it is assumed for beginning that this vector is the same for all defects), $J_{\bf k} = 2J( \cos k_x + \cos k_z )$, where $J$ is the exchange constant, and $\epsilon_{\bf k} = s\sqrt{J_0^2-J_{\bf k}^2}$.

Let us discuss the spin-wave spectrum near points $k = 0$ and $k = k_0$, where ${\bf k}_0$ is the antiferromagnetic vector. It is seen that expression (\ref{sgfr}) is symmetric under replacement of $\bf k$ by ${\bf k} \pm {\bf k}_0$. Thus, we consider below only the vicinity of the point $k=0$. We have from Eq.~(\ref{sgfr}):
\begin{eqnarray}
\label{fg2}
{\cal D}(\omega,{\bf k}) &=& \omega^2 - \epsilon_{\bf k}^2 \left[ 1 - \frac{n f^2}{2\pi} \Theta u({\bf k})\chi_\perp(\omega) \right],\\
\label{c}
u({\bf k}) &=& \frac 12 + \frac{({\bf kR}_{12})^2}{k^2},
\end{eqnarray}
where it is used that the unperturbed spectrum is linear at $k\ll k_0$: $\epsilon_{\bf k}=ck=\sqrt8 sJk$. It is seen from Eqs.~(\ref{fg2}) and (\ref{c}) that the spectrum appears to be dependent on the direction of the momentum $\bf k$ as a result of interaction of magnons with the defects. This circumstance is a consequence of our assumption that the vector ${\bf R}_{12}$ is the same for all impurities. In fact, it can have four directions and the value $({\bf R}_{12}{\bf k})^2/k^2$ can have two different values: $\cos^2 \phi_{\bf k}$ and  $\sin^2 \phi_{\bf k}$, where $\phi_{\bf k}$ is the azimuthal angle of $\bf k$. It easy to realize that $u({\bf k})=1$ if all four ways of coupling of the impurity with AF are equally possible.

It is convenient for the following to consider separately the cases of $|\omega| \gg \omega_0$ and $|\omega| \ll \omega_0$, where $\omega_0$ is given by Eq.~(\ref{omlim}). In these cases the non-resonant and the resonant parts, respectively, are dominant in the impurity susceptibility (\ref{chi3}).

\underline{$|\omega| \gg \omega_0$}. One obtains from Eqs.~(\ref{chi3}), (\ref{imchi}) and (\ref{fg2}) for the magnon damping $\gamma_{\bf k}$ and the renormalized spin-wave velocity $\tilde c_{\bf k}$:
\begin{eqnarray}
\label{damping}
\gamma_{\bf k} &=& |\omega|\frac{n f^4}{16\pi} u({\bf k}) ,\\
\label{veloc}
\tilde c_{\bf k}^2 &=& c^2 \left( 1 - \ln\left|\frac{\Theta}{\omega}\right| \frac{n f^4}{4\pi^2} u({\bf k}) \right),
\end{eqnarray}
where we take into account that $R_\chi = 1/2$. It is seen that the interaction with the defects leads to strong damping which is proportional to $\omega$ and to the logarithmic correction to the spin-wave velocity. It would seem that at small enough $k$ the spin-wave velocity becomes imaginary signifying a phase transition in the system. Meanwhile our theory is not applicable at such small $k$. The interaction with the defects changes the function ${\rm Im} \Delta_{\mu\nu}(\omega)$ as well and this renormalization is strong at small enough $\omega$. Indeed, one has to use renormalized spin Green's function derived in Appendix~\ref{gfwi} to evaluate ${\rm Im}\Delta_{\mu\nu}(\omega)$. As a result of simple calculations similar to those presented in Appendix~\ref{delprop} we obtain that at $|\omega|\ll\Theta$ in addition to the term proportional to $\omega^2{\rm sgn}(\omega)$ there is another one proportional to ${\rm sgn}(\omega)$:
\begin{eqnarray}
\label{delad1}
{\rm Im} \Delta_{\mu\nu}(\omega) &=& -A \left(\frac{\omega}{\Theta}\right)^2 {\rm sgn}(\omega)d_{\mu\nu} 
-
{\rm sgn}(\omega)\frac{snf^4B}{\Theta}d_{\mu\nu},\\
\label{b}
B &=& 
\frac{s^4J^2}{\pi^{5/2}}
\int  
d{\bf k}
\frac{(1+\cos({\bf k R}_{12}))^2(J_{\bf 0} - J_{\bf k})^2}{\epsilon_{\bf k}^4}
\approx 0.2,
\end{eqnarray}
where the integral is taken over the chemical Brillouin zone. The first term in Eq.~(\ref{delad1}) is greater than the second one when
\begin{equation}
\label{o1}
|\omega| \gg 0.02\Theta \sqrt n f^2.
\end{equation}
This condition determines the range of validity of our theory at $|\omega| \gg \omega_0$. At $\omega$ given by (\ref{o1}) the logarithmic correction to the spin-wave velocity in Eq.~(\ref{veloc}) is small.

\underline{$|\omega| \ll \omega_0$}. In this case the impurity susceptibility (\ref{chi3}) is determined by the resonant term. We have for the spin-wave damping and the spin-wave velocity from Eqs.~(\ref{chi3}), (\ref{imchi}) and (\ref{fg2}):
\begin{eqnarray}
\label{damp2}
\gamma_{\bf k} &=& \frac\Theta T \frac{\omega^2 \varGamma}{\omega^2 + 4\varGamma^2} \frac{n f^2}{8\pi}u({\bf k}),\\
\label{vel2}
\tilde c_{\bf k}^2 &=& c^2 \left( 1 - \frac\Theta T \frac{\varGamma^2}{\omega^2 + 4\varGamma^2} \frac{n f^2}{2\pi}u({\bf k}) \right).
\end{eqnarray}
As in the case of $|\omega| \gg \omega_0$, one has to take into account the renormalization of the function ${\rm Im} \Delta_{\mu\nu}(\omega)$. After simple calculations we obtain at $|\omega|\ll\Theta$:
\begin{equation}
\label{delad2}
{\rm Im} \Delta_{\mu\nu}(\omega) = -A \left(\frac{\omega}{\Theta}\right)^2 {\rm sgn}(\omega)d_{\mu\nu} 
- 2snf^2B \frac{\omega\varGamma}{T(\omega^2 + 4\varGamma^2)} d_{\mu\nu},
\end{equation}
where the constant $B$ is given by Eq.~(\ref{b}). As a result the range of validity of our consideration is determined by
\begin{equation}
\label{ol}
\frac{|\omega|(\omega^2+4\varGamma^2)}{\Theta^3} \gg 0.004nf^2\frac{\varGamma}{T}.
\end{equation}
It is easy to show that the spin-wave damping (\ref{damp2}) and the correction to the spin-wave velocity in Eq.~(\ref{vel2}) is small at such $\omega$.

It should be noted once more that if one of the conditions (\ref{o1}) or (\ref{ol}) is violated the interaction between defects becomes important and our approach is wrong.

\subsection{Specific heat}

\begin{figure}
\centering
\includegraphics[scale=0.9]{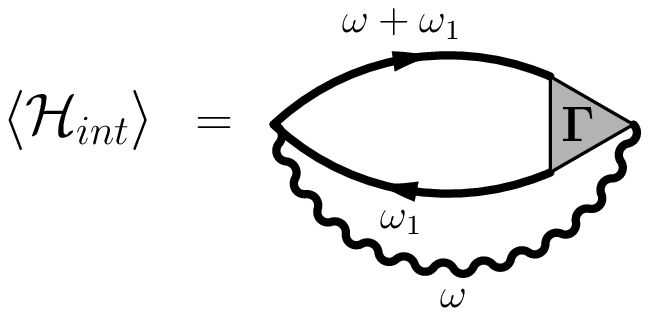}
\caption{Graphical representation of averaged interaction of one impurity with 2D AF. Bold solid and bold wavy lines denote, respectively, dressed pseudofermion Green's functions and the boson Green's function renormalized by interaction with impurities.
\label{heatf}} 
\end{figure}

We proceed with the discussion of the magnetic part of the specific heat $C(T)$. It is convenient to use the following formula for its evaluation:
\begin{equation}
\label{heat}
C(T) = \frac{dE}{dT} = \frac{d}{dT}\left\langle \sum_{\bf k}\epsilon_{\bf k}\left(\alpha^\dagger_{\bf k}\alpha_{\bf k}+\frac12\right) + \sum_i{\cal H}^{(i)}_{int}\right\rangle,
\end{equation}
where the first term describes magnetic excitations and index $i$ in the second term labels impurities. To calculate the first term in Eq.~(\ref{heat}) we use bilinear part of the Hamiltonian (\ref{h2}) and spin Green's functions (\ref{gfi}) renormalized by interaction with impurities. As a result of simple calculations one obtains up to inessential terms not depending on the temperature:
\begin{eqnarray}
\label{h0av}
\left\langle \sum_{\bf k}\epsilon_{\bf k}\left(\alpha^\dagger_{\bf k}\alpha_{\bf k}+\frac12\right)\right\rangle 
&=&
\sum_{\bf k} \epsilon_{\bf k} N(\epsilon_{\bf k})
+
2Nnf^2\Theta X\int_{-\infty}^\infty d\omega N(\omega) {\rm Im}\chi_\perp(\omega),\\
\label{xconst}
X &=& 
\frac{s^2J}{2\pi^4}
\int  
d{\bf k}
\frac{J_{\bf 0} - J_{\bf k}\cos({\bf k R}_{12})}{\epsilon_{\bf k}^2}
\approx 0.05,
\end{eqnarray}
where the integration in Eq.~(\ref{xconst}) is over the chemical Brillouin zone. The diagram for the second term in Eq.~(\ref{heat}) is shown in Fig.~\ref{heatf}. It contains the impurity susceptibility calculated above and $\Delta_{\mu\nu}(\omega)$:
\begin{equation}
\label{diaint}
\left\langle \sum_i {\cal H}_{int}^{(i)} \right\rangle = N\frac{ng^2}{\pi}\sum_\nu\int_{-\infty}^\infty d\omega N(\omega) [{\rm Im}\chi_\nu(\omega){\rm Re}\Delta_{\nu\nu}(\omega) + {\rm Re}\chi_\nu(\omega){\rm Im}\Delta_{\nu\nu}(\omega)].
\end{equation}
Evaluating expression (\ref{diaint}) and summing it with Eq.~(\ref{h0av}) one leads to the following expression:
\begin{eqnarray}
\label{avh}
\frac EN &=& 
\frac1N\sum_{\bf k} \epsilon_{\bf k} N(\epsilon_{\bf k})
-
\frac{2nf^2}{\pi\Theta}\int_{-\infty}^\infty d\omega N(\omega)\omega|\omega|\Lambda(\omega) {\rm Re}\chi_\perp(\omega)
+
nf^2\Theta X\int_{-\infty}^\infty d\omega N(\omega) {\rm Im}\chi_\perp(\omega),
\end{eqnarray}
where the constant $X$ is given by Eq.~(\ref{xconst}). The first term in Eq.~(\ref{avh}) describes the energy of 2D AF without impurities. It is equal to $E_0 \propto NT^3/\Theta^2$ at $T\ll\Theta$ and we come to the well known result: $C(T)\propto T^2$ for pure 2D AF. Evaluation of corrections to $E_0$ from the second and the third terms in Eq.~(\ref{avh}) is a tedious but straightforward work. Unfortunately Eq.~(\ref{chi3}) for $\chi_\perp(\omega)$ has the limited range of validity which is discussed in the previous section. Therefore one can not carry out the integration in the second and the third terms of Eq.~(\ref{avh}) in the whole range of $\omega$. We have evaluated these terms after the integration over $\omega$ at which Eq.~(\ref{chi3}) is valid. There are corrections having weaker $T$-dependence than $E_0$ stemming from both the resonant and the nonresonant parts of $\chi_\perp(\omega)$. Meanwhile restrictions (\ref{o1}) and (\ref{ol}) result in the considered corrections to be bounded below on $T$ and to be smaller than $E_0$. 

Thus, we do not obtain a renormalization of the specific heat within our precision.

\section{Conclusion}
\label{concl}

We discuss the dynamical properties of the impurity spin-$\frac12$ in 2D and quasi-2D Heisenberg antiferromagnets (AFs) at $T\ge0$. The specific case of the impurity that is coupled {\it symmetrically} to two neighboring host spins is considered. It is shown that this problem is a generalization of the spin-boson model without the tunneling term and with a more complex interaction. It is demonstrated that the effect of the host system on the defect is completely described by the spectral function ${\cal J}(\omega)$ which is proportional to $\omega^2$. It is found within the spin-wave approximation that the spectral function has this $\omega$-dependence in 2D AF for not too small $\omega$. In isotropic 2D AF at $T>0$ ${\cal J}(\omega)\propto\omega^2$ when $\omega\gg Ja/\xi$, where $J$ is the exchange constant between the host spins, $\xi$ is the correlation length and $a$ is the lattice constant. For ordered 2D AF ${\cal J}(\omega)\propto\omega^2$ at $\omega\gg\eta$, where $\eta\ll J$ is the value of interaction (for definiteness interplane interaction) stabilizing the long range order at finite $T$.

We stress that one must distinguish symmetrically and asymmetrically coupled impurities (see Fig.~\ref{pic}). Symmetrically coupled impurity is located in the zero molecular field. It remains degenerate and the spectral function is proportional to $\omega^2$. In the case of asymmetrically coupled impurity, where the molecular field is nonzero, there is splitting of the impurity levels and the spectral function has terms with weaker $\omega$-dependence. For instance, we demonstrate that the spectral function for impurity coupled to one host spin is proportional to a constant. In this paper we consider only the symmetric case. Our results are also valid with certain additional restrictions for slightly split nearly symmetrically coupled impurities (see below).

The defect dynamical susceptibility $\chi(\omega)$ is derived using Abrikosov's pseudofermion technique and diagrammatic expansion. The calculations are performed within the order of $f^4$, where $f\propto g/J$ is the dimensionless coupling parameter. For our study the sign of $f$ is insignificant. We show that the transverse impurity susceptibility $\chi_\perp(\omega)$ has a Lorenz peak with the widths $\varGamma\propto f^4J(T/J)^3$ that disappears at $T=0$, and a non-resonant term. The longitudinal susceptibility $\chi_\|(\omega)$ has the non-resonant term which differs from that of $\chi_\perp(\omega)$ by a constant and a Lorenz peak. The width of the peak is zero within the order of $f^4$. Its calculation is out of the scope of this paper. The imaginary part of non-resonant term is a constant independent of $T$ at $|\omega|\gg\varGamma$ and the real part has a logarithmic divergence as $\omega,T\to0$. Similar logarithmic singularity was found in Ref.~\cite{nagaosa} at $T=0$. 

The static susceptibility has the free-spin-like term $S(S+1)/(3T)$ and a correction proportional to $f^2\ln (J/T)$. We point out here the sharp difference between symmetrically and asymmetrically coupled impurities that takes place in the regime $T\ll |g|$ (by asymmetrically coupled impurities we mean here either the added spin coupled to one host spin or the vacancy which is the particular case of the added spin with $g\to\infty$). The leading $1/T$-term has the free-spin-like form in the symmetric case and the classical-like form in the asymmetric one. Moreover, the logarithmic correction is proportional to $g^2$ in the symmetric case and it does not depend on $g$ in the asymmetric one. \cite{sush2,sachdev} The difference is related to the fact that the impurity spin coupled asymmetrically aligns with the local Neel order, \cite{sush2,sachdev} whereas the symmetrically coupled impurity is located in the zero molecular field.

The fact that the spectral function in 2D AF is proportional to $\omega^2$ only at $\omega\gg\{\eta \mbox{ \rm or } Ja/\xi\}$ leads to the following restriction on the results obtained: $\max\{\varGamma_0,|\omega|\}\gg\{\eta \mbox{ \rm or } Ja/\xi\}$. If the defect is slightly split (for definiteness by magnetic field $\bf H$) this condition turns into $\max\{\varGamma_0,|\omega|\}\gg\max\{\{\eta \mbox{ \rm or } Ja/\xi\},g\mu_BHS\}$. For nearly symmetrically coupled impurity one has: $\max\{\varGamma_0,|\omega|\}\gg\max\{\{\eta \mbox{ \rm or } Ja/\xi\},|g_1-g_2|\}$, where $g_{1,2}$ are values of coupling with the corresponding sublattices (see Fig.~\ref{pic}).

The findings discussed above are valid for isotropic interaction of the impurity spin $\bf S$ with AF: ${\cal H}_{int} = g{\bf S}({\bf s}_1 + {\bf s}_2)$, where ${\bf s}_{1,2}$ are host spins from different sublattices. We consider also interaction containing only one component of $\bf S$: ${\cal H}_{int} = gS^x(s_1^x+s_2^x)$. The results in this case are quite specific. We show that $xx$- component of the impurity susceptibility is zero whereas $yy$- and $zz$- ones have only the non-resonant term. Our model is identical to the spin-boson one (\ref{ham1}) without ${\cal H}_d$ if the interaction contains a term with only one component of $\bf S$. The Hamiltonian can be diagonalized exactly \cite{pirc} and an exact expression for $\chi(\omega)$ can be obtained. We perform the corresponding calculations and confirm the results obtained by our approach. One of the most interesting features of the exact result is that the static susceptibility has the form $\chi(0) \propto T^{-1-\zeta}$, where $\zeta\propto f^2T/J$. Within the first order of $f^2$ one has $1/(4T)$-term and the logarithmic correction. Thus, we see that in the modified spin-boson model taking into account the higher order logarithmic corrections results in the non-trivial power-law $T$-dependence of $\chi(0)$.

The influence of the finite concentration $n$ of the defects on the low-temperature properties of 2D AF is also considered. For not too small $\omega$ we find the logarithmic correction to the spin-wave velocity of the form $nf^4\ln|J/\omega|$ and an anomalous damping of the spin-waves proportional to $nf^4|\omega|$. Similar logarithmic correction to the velocity and damping were obtained in Ref.~\cite{chern}, where vacancies in 2D AF were studied. It is demonstrated that interaction of the spin waves with defects modifies the spectral function which acquires new terms proportional to $n$ exhibiting weaker $\omega$-dependence. These terms should be taken into account at small enough $\omega$ and the problem should be solved self-consistently. The corresponding consideration is out of the scope of this paper. Within the range of validity of our study we do not obtain a renormalization of the magnetic specific heat which is proportional to $T^2$ in 2D AF without impurities.

The results of the present paper can be applied to other systems with a degenerate defect in which the spectral function is proportional to $\omega^2$. 

\begin{acknowledgments}
We are thankful to A. V. Lazuta for stimulating discussions. This work was supported by Russian Science Support Foundation (A.V.S.), RFBR (Grant Nos. SS-1671.2003.2, 03-02-17340, 06-02-16702 and 00-15-96814), and Russian Programs "Quantum Macrophysics", "Strongly correlated electrons in semiconductors, metals, superconductors and magnetic materials" and "Neutron Research of Solids".
\end{acknowledgments}

\appendix

\section{Calculation of ${\rm Im} \Delta_{\mu\nu}(\omega)$ in 2D AF}
\label{delprop}

In this appendix we discuss properties of the imaginary part of the function $\Delta_{\mu\nu}(\omega)$ general expression for which is given by Eq.~(\ref{deltaf}). It is shown below that within the spin-wave approximation ${\rm Im}\Delta_{\mu\nu}(\omega)$ has the form (\ref{delta}) and expressions for the constant $A$, the characteristic energy $\Theta$ and the tensor $d_{\mu\nu}$ are obtained.

We have from Eqs.~(\ref{deltaf}) and (\ref{intaf}):
\begin{equation}
\label{delta0}
\Delta_{\mu\nu}(\omega) = \frac2N \sum_{\bf k} [1+\cos({\bf k R}_{12})] \langle s^\mu_{-\bf k} s^\nu_{\bf k} \rangle_\omega, 
\end{equation}
where $\langle \dots \rangle_\omega$ denote retarded Green's function, $N$ is the number of spins in the lattice and ${\bf R}_{12}$ is the vector connecting two host spins coupled to the defect. Thus we have to calculate the spin Green's functions $\langle s^\mu_{-\bf k} s^\nu_{\bf k} \rangle_\omega$ of 2D AF. The Hamiltonian of 2D Heisenberg AF on the square lattice has the well-known form:
\begin{equation}
\label{ham2d}
H = J\sum_{\langle ij \rangle} {\bf s}_i {\bf s}_j.
\end{equation}
We perform for beginning all calculations for isotropic Heisenberg 2D AF at $T=0$ and then consider the effect of finite $T$ and of the additional small interaction stabilizing the long range order at finite $T$. 

Instead of dividing of the lattice onto two sublattices it is convenient to represent operators $s^\mu_{\bf k}$ as follows (see, e.g., Refs.~\cite{malold,pet}):
\begin{equation}
\label{sk}
{\bf s}_{\bf k} = \hat x s^x_{\bf k} + \hat y s^y_{{\bf k} + {\bf k}_0} + \hat z s^z_{{\bf k} + {\bf k}_0},
\end{equation}
\begin{equation}
\label{srep}
s^x_{\bf k} = \sqrt{\frac s2} \left( a_{\bf k} + a^\dagger_{-\bf k} - \frac{(a^2a^\dagger)_{\bf k}}{2s} \right), \qquad 
s^y_{\bf k} = -i\sqrt{\frac s2} \left( a_{\bf k} - a^\dagger_{-\bf k} - \frac{(a^2a^\dagger)_{\bf k}}{2s} \right), \qquad  
s^z_{\bf k} = s - (a^\dagger a)_{\bf k},
\end{equation}
where $z$ axis is parallel to the magnetization of sublattices, $ \hat x, \hat y, \hat z$ denote unit vectors directed along corresponding axes, ${\bf k}_0 = (\pi,0,\pi)$ is the antiferromagnetic vector and $s$ is the spin value. Substitution of Eqs.~(\ref{sk}) and (\ref{srep}) to Eq.~(\ref{ham2d}) leads to the following expression for the Hamiltonian: $H = E_0 + \sum_{i=2}^6 H_i$, where $E_0$ is the ground state energy and $H_i$ denote terms containing products of $i$ operators $a$ and $a^\dagger$. We consider in this paper the spin-wave approximation, i.e., we restrict ourself by the bilinear part of the Hamiltonian which has the form
\begin{equation}
\label{h2}
H_2 = \sum_{\bf k} \left[E_{\bf k} a^\dagger_{\bf k}a_{\bf k} + 
\frac{B_{\bf k}}{2} \left( a^\dagger_{\bf k}a^\dagger_{-\bf k} + a_{\bf k}a_{-\bf k} \right)\right],
\end{equation}
where $E_{\bf k} = sJ_0$, $B_{\bf k} = sJ_{\bf k}$ and $J_{\bf k} = 2J( \cos k_x + \cos k_z )$. As is seen from Eqs.~(\ref{delta0}), (\ref{sk}) and (\ref{srep}), the only nonzero components of the spin Green's function are $xx$ and $yy$ and tensor $d_{\mu\nu}$ has the form:
\begin{equation}
\label{daf}
d_{\mu\nu} = 
\left\{
\begin{array}{ll}
\delta_{\mu\nu}, \quad &  \mbox{ if } \mu,\nu = x,y,\\
0, &  \mbox{ if }\mu = z  \mbox{ or } \nu = z.
\end{array}
\right.
\end{equation}
The corresponding components of $\Delta_{\mu\nu}(\omega)$ can be derived using Green's functions $g(\omega,{\bf k}) = \langle a_{\bf k}, a^\dagger_{\bf k} \rangle_\omega$, $f(\omega,{\bf k}) = \langle a_{\bf k}, a_{-\bf k} \rangle_\omega$, $\bar g(\omega,{\bf k}) = \langle a^\dagger_{-\bf k}, a_{-\bf k} \rangle_\omega = g^*(-\omega,-{\bf k})$ and $f^\dagger (\omega,{\bf k}) = \langle a^\dagger_{-\bf k}, a^\dagger_{\bf k} \rangle_\omega = f^*(-\omega,-{\bf k})$. For two of them we have the Dyson equation:
\begin{equation}
\label{eqfunc}
\begin{array}{l}
g(\omega,{\bf k}) = g^{(0)}(\omega,{\bf k}) + g^{(0)}(\omega,{\bf k}) B_{\bf k} f^\dagger(\omega,{\bf k}),\\
f^\dagger(\omega,{\bf k}) = \bar g^{(0)}(\omega,{\bf k}) B_{\bf k} g(\omega,{\bf k}),
\end{array}
\end{equation}
where $g^{(0)}(\omega,{\bf k}) = (\omega - E_{\bf k}+i\delta)^{-1}$ is the bare Green's function. Solving Eq.~(\ref{eqfunc}) one obtains:
\begin{equation}
\label{gf}
g(\omega,{\bf k}) = \frac{\omega + sJ_{\bf 0}}{(\omega + i\delta)^2 - \epsilon_{\bf k}^2},
\quad
f(\omega,{\bf k}) = -\frac{sJ_{\bf k}}{(\omega + i\delta)^2 - \epsilon_{\bf k}^2},
\end{equation}
where $\epsilon_{\bf k} = s\sqrt{J_0^2-J_{\bf k}^2}$ is the spin-wave energy. As a result of direct calculations we have:
\begin{equation}
\label{delr}
{\rm Im} \Delta_{\mu\nu}(\omega) = -d_{\mu\nu}\frac{s^2}{4\pi}\int d{\bf k} \left[\{1+\cos({\bf k R}_{12})\}(J_{\bf 0} - J_{\bf k}) + \{1 + \cos(({\bf k} + {\bf k}_0) {\bf R}_{12})\}(J_{\bf 0} + J_{\bf k}) \right] 
\frac{1}{\epsilon_{\bf k}}
\left[\delta(\omega-\epsilon_{\bf k}) - \delta(\omega+\epsilon_{\bf k}) \right],
\end{equation}
where the lattice constant is taken to be equal to unity and the integral is over the magnetic Brillouin zone. If $|\omega|\ll sJ$ we have $J_{\bf k}\approx J_{\bf 0}(1 - k^2/4)$, $\epsilon_{\bf k}=ck=\sqrt8 sJk$ and $\cos({\bf kR}_{12})\approx 1-({\bf kR}_{12})^2/2$. Notice that $({\bf k}_0 {\bf R}_{12}) = \pi \mod 2\pi$ if the impurity is coupled to spins from different sublattices and both terms in the first square brackets in Eq.~(\ref{delr}) are proportional to $k^2$. Then integration in Eq.~(\ref{delr}) can be easily carried out if one takes advantage of the approximation for magnons similar to Debye one for phonons: the spectrum is assumed to be linear, $\epsilon_{\bf k}=ck$, up to cut-off momentum $k_\Theta$ defined from the equation $2N = V(2\pi)^{-1}\int_0^{k_\Theta}dkk$, where $V$ is the area of the lattice. As a result we lead to expression (\ref{delta}) for ${\rm Im} \Delta_{\mu\nu}(\omega)$, where 
\begin{equation}
\label{constants}
\Theta = ck_\Theta = 8\sqrt\pi sJ, \quad A=\frac{2 \pi}{J}.
\end{equation}
The factor $A$ should be multiplied by 2 if the defect is coupled to four host spins (two by two from each sublattice). 

It is well known that there is no long range order in Heisenberg 2D AF at $T>0$. \cite{mw} Nevertheless it is shown theoretically \cite{kop,chak,tyc} and confirmed experimentally \cite{thur} that the spin waves are well defined in paramagnetic phase of 2D AF if their wavelength is much smaller than the correlation length $\xi\propto \exp({\rm const}/T)$. Thus, the above result for ${\rm Im}\Delta_{\mu\nu}(\omega)$ is valid when $|\omega|\gg Ja/\xi$, where $a$ is the lattice spacing.

It is easy to conclude that if a small interplane interaction of the value of $\eta\ll J$ is taken into account the above result for ${\rm Im}\Delta_{\mu\nu}(\omega)$ is valid when $|\omega|\gg\eta$ (see discussion in Sec.~\ref{model}). At the same time ${\rm Im}\Delta_{\mu\nu}(\omega)$ has another $\omega$-dependence if $|\omega|\alt\eta$. 

Finally, we note that when the impurity is coupled to one host spin we have $\Delta_{\mu\nu}(\omega) = N^{-1} \sum_{\bf k} \langle s^\mu_{-\bf k} s^\nu_{\bf k} \rangle_\omega$ instead of Eq.~(\ref{delta0}). Comparing this equation with (\ref{delta0}) and (\ref{delr}) one infers that the spectral function is proportional to a constant in this case.

\section{Matrix structure of pseudofermion Green's function and the vertex for 2D AF}
\label{diag}

In this appendix we discuss the matrix structure of the dressed pseudofermion Green's functions $G_{mm'}(x)$ and the pseudofermion vertex $\Gamma_{Pmm'}(x+\omega,x)$ for 2D AF. Some lower-order diagrams for the self-energy $\Sigma_{mm'}(\omega)$ and for $\Gamma_{Pmm'}$ are shown in Figs.~\ref{sigmafig} and \ref{gammaf}, respectively. Let us discuss firstly the self-energy. Its matrix structure is determined by the corresponding products of operators $S^\mu$ and tensors $d_{\mu\nu}$. For instance, this product for the second diagram in Fig.~\ref{sigmafig} has the form $\sum_{\mu\mu'\nu\nu'} S^\mu S^\nu S^{\mu'} S^{\nu'} d_{\nu\nu'}d_{\mu\mu'}$. One can make sure that such combinations are proportional to the unit matrix using the following evident representation of an arbitrary matrix $A$ of the size $2\times2$ via Pauli matrices:
\begin{equation}
\label{ex}
A = a_0 + (\mbox{\boldmath $\sigma$}{\bf a}),
\end{equation}
where $a_0 = {\rm Tr}(A)/2$ and ${\bf a} = {\rm Tr}(\mbox{\boldmath $\sigma$} A)/2$. It is seen from the view of the tensor $d_{\mu\nu}$ given by Eq.~(\ref{daf}) that the combinations of $S^\mu$ contain products of even number of matrices $\sigma_x$ and $\sigma_y$. According to (\ref{ex}) such combinations are proportional to the unit matrix.

The similar consideration can be carried out for the vertex $\Gamma_{Pmm'}(x+\omega,x)$, where $P$ is one of Pauli matrices. The vertex matrix structure is determined by the products of operators $S^\mu$, tensors $d_{\mu\nu}$ and $P$. For example, the corresponding product for the fourth diagram in Fig.~\ref{gammaf} has the form $\sum_{\mu\mu'\nu\nu'}S^\mu S^\nu P S^{\mu'} S^{\nu'}d_{\nu\nu'}d_{\mu\mu'}$. It can be easily shown using (\ref{ex}) that such combinations are proportional to $P$.

\section{Calculation of the impurity susceptibility}
\label{gameval}

We present in this appendix some details of the impurity dynamical susceptibility calculation. We use for this the general expression (\ref{chi}) and Eqs.~(\ref{2gf}), (\ref{2g++}) and (\ref{2g+-2}) for the Green's function and the branches of the vertex. The expression for $\chi_P(\omega)$ is derived up to the order of $f^2$. In this order the interaction does not change the average number of pseudofermions $\cal N$ given by Eq.~(\ref{n}), i.e. ${\cal N} = 2e^{-\lambda/T}$. To show this let us express $\cal N$ as an integral of the Green's function: \cite{agd}
\begin{equation}
\label{n2}
{\cal N} = -\frac1\pi\sum_m\int_{-\infty}^\infty dx n(x){\rm Im}G_{mm}(x) 
= -\frac{2e^{-\lambda/T}}{\pi} \int_{-\infty}^\infty dx e^{-x/T}{\rm Im}G(x)
\end{equation}
where $n(x) = (e^{x/T} + 1)^{-1}$ is the Fermi function and $G(x)$ is given by Eq.~(\ref{2gf}). We make a shift by $\lambda$ in the last part of Eq.~(\ref{n2}) and replace $n(x+\lambda)$ by $e^{-(x+\lambda)/T}$ as it was done in Eq.~(\ref{chi}) for $\chi_P(\omega)$. Because $\gamma(x)$ and $Z(x)$ are exponentially small at negative $x$ if $|x|\gg T$ (see Eq.~(\ref{prop})) the integrand in Eq.~(\ref{n2}) does not increase exponentially as $x\to-\infty$. It is easy to make sure that the terms proportional to $f^2$ cancel each other in Eq.~(\ref{n2}).

According to Eqs.~(\ref{chi}), (\ref{2g++}) and (\ref{gam+-tls}) the dynamical susceptibility can be represented as a sum of three components. The first one, $\chi_1(\omega)$, originates from Eq.~(\ref{chi}) as a result of replacement of the vertex by unity. The second, $\chi_2(\omega)$, appears from $f^2$-terms in Eqs.~(\ref{2g++}) and (\ref{gam+-tls}). The third, $\chi_3(\omega)$, is a result of replacement of the vertex by the third term from Eq.~(\ref{gam+-tls}).

The expression for $\chi_1(\omega)$ can be brought to the form
\begin{equation}
\label{chi11}
\chi_1(\omega) = \frac{\overline{P^2}}{2\pi}\int_{-\infty}^\infty dx e^{-x/T} [G(x+\omega) + G^*(x-\omega)] {\rm Im} G(x).
\end{equation}
The integrand in Eq.~(\ref{chi11}) does not increase exponentially as $x \to -\infty$ because $\gamma(x)$ and $Z(x)$ obey the property (\ref{prop}). Using Eq.~(\ref{2gf}) for Green's functions it is convenient to represent Eq.~(\ref{chi11}) in the following form:
\begin{eqnarray}
\label{chi12}
\chi_1(\omega) &=& [ J_0(\omega) + J_1(\omega) + J_2(\omega) ] + [ J^*_0(-\omega) + J^*_1(-\omega) + J^*_2(-\omega) ],\\
\label{j01}
J_0(\omega) &=& -\frac{\overline{P^2} }{2\pi} \int_{-\infty}^\infty dx e^{-x/T} \frac{1}{x+\omega+i\varGamma_0} \frac{\gamma(x)}{x^2 + \varGamma_0^2},\\
\label{j11}
J_1(\omega) &=& -\frac{\overline{P^2} }{4\pi i} \int_{-\infty}^\infty dx e^{-x/T} \frac{1}{x+\omega+i\varGamma_0} \left[
\frac{-i\gamma(x) [Z(x)+Z^*(x)] + x [Z(x)-Z^*(x)]}{x^2 + \varGamma_0^2}
\right],\\
\label{j21}
J_2(\omega) &=& \frac{\overline{P^2} }{2\pi} \int_{-\infty}^\infty dx e^{-x/T} \frac{Z(x+\omega)}{x+\omega+i\varGamma_0} \frac{\gamma(x)}{x^2 + \varGamma_0^2},
\end{eqnarray}
where in all denominators we replace $\gamma(x)$ and $\gamma(x+\omega)$ by $\varGamma_0$, their values at $|x| \ll T$ and  $|x+\omega| \ll T$, respectively (see Eq.~(\ref{2gamm0})). It can be done because $|\gamma(x)|\ll |x|$ at $|x|\agt T$.

The integration in Eq.~(\ref{j01}) can be easily carried out if one notes that the main contribution arises from the area of $x\sim \omega,\varGamma$, where $e^{-x/T} \approx 1-x/T$ (it will be clear soon that the second term in this expansion is essential). As a result we have:
\begin{equation}
\label{j02}
J_0(\omega) = -\frac{\overline{P^2}}{2}\frac{1}{\omega+2i\varGamma_0}\left(1-\frac{i\varGamma_0}{T}\right).
\end{equation}

\begin{figure}
\centering
\includegraphics[scale=0.7]{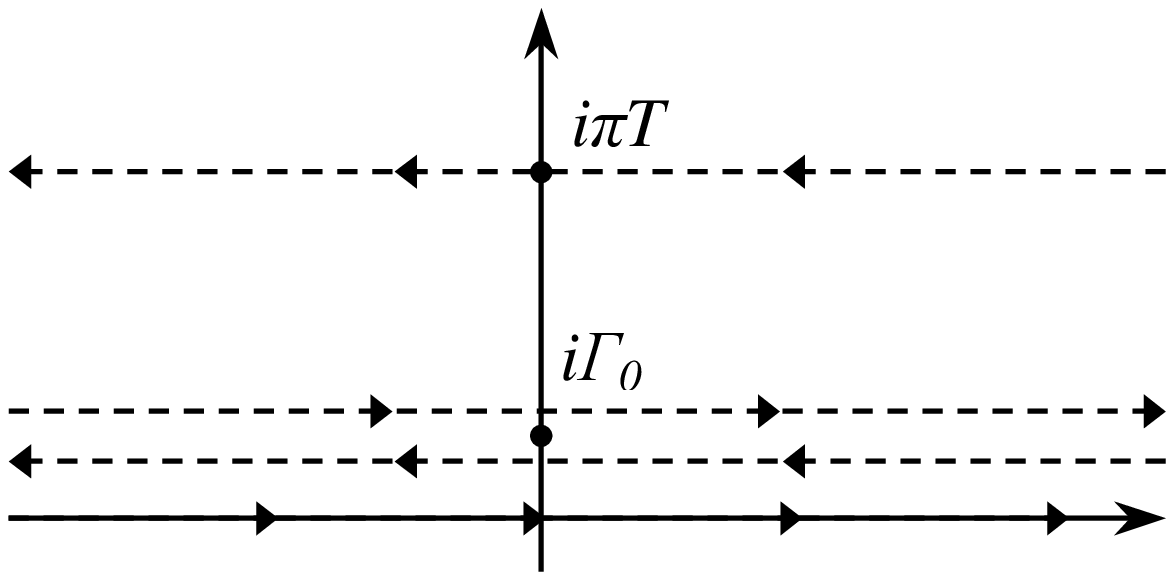}
\caption{Contour of integration used in Appendix~\ref{gameval} for the impurity susceptibility calculation. This contour consists of four lines which are parallel to the real axis. They pass through points $x=0$, $x=i\varGamma_0-i\delta$, $x=i\varGamma_0+i\delta$ and $x=i\pi T$.
\label{contour}} 
\end{figure}

To take the integral in Eq.~(\ref{j11}) for $J_1(\omega)$ we consider a contour integral with the same integrand. The contour is presented in Fig.~\ref{contour}. It consists of four lines which are parallel to the real axis. They pass through points $x=0$, $x=i\varGamma_0-i\delta$, $x=i\varGamma_0+i\delta$ and $x=i\pi T$. It can be shown using definitions of $\gamma(x)$ and $Z(x)$ that the integrand is an analytical function inside the contour. Note, the contour envelops the cut of $Z(x)^*$ passing through the point $x=i\varGamma_0$ along the real axis. As a result we have:
\begin{eqnarray}
\label{j12}
J_1(\omega) &=& -\frac{\overline{P^2}}{2\pi}\frac{f^2R_\Sigma}{\pi\Theta}
\int_{-\infty}^\infty dx dx_1 e^{-x/T}|x_1|N(x_1)\Lambda(x_1)
\frac{\gamma(x+i\pi T)(x_1+x+i\pi T) + \gamma(x_1+x+i\pi T)(x+i\pi T)}{(x+i\pi T)^3 (x_1+x+i\pi T)^2}\nonumber\\
&&{}+
\frac{\overline{P^2}}{4\pi i}\frac{f^2R_\Sigma}{\pi\Theta}e^{-i\varGamma_0/T}
\int_{-\infty}^\infty dx dx_1 \frac{e^{-x/T}|x_1|N(x_1)\Lambda(x_1)}{x+\omega+2i\varGamma_0}
\left[
\frac{1}{x_1+x-i\delta}\frac{1}{x-i\delta} - \frac{1}{x_1+x+i\delta}\frac{1}{x+i\delta}
\right],
\end{eqnarray}
where the first term is the result of integration over the line of the contour passing through $x=i\pi T$ and the second term describes the sum of integration over lines passing through $x=i\varGamma_0-i\delta$ and $x=i\varGamma_0+i\delta$. It is seen from Eq.~(\ref{j12}) that the first term is of the order of $f^6$ and can be discarded. The second term in Eq.~(\ref{j12}) can be simply calculated using the equation $(x_1+x-i\delta)^{-1}(x-i\delta)^{-1} - (x_1+x+i\delta)^{-1}(x+i\delta)^{-1} = 2\pi i [\delta(x_1+x)(x-i\delta)^{-1} + \delta(x)(x_1+i\delta)^{-1}]$. As a result we have:
\begin{equation}
\label{j13}
J_1(\omega) = \frac{\overline{P^2}}{2} \frac{e^{-i\varGamma_0/T}}{\omega+2i\varGamma_0}\frac{f^2R_\Sigma}{\pi\Theta}
\int_{-\infty}^\infty dx\frac{|x|N(x)\Lambda(x)}{x+\omega+2i\varGamma_0}.
\end{equation}
One can carry out the integration in Eq.~(\ref{j21}) for $J_2(\omega)$ in the similar way. As a result we obtain that $J_2(\omega) = J_1(\omega)$ and $\chi_1(\omega)$ has the form:
\begin{equation}
\label{chi1}
\chi_1(\omega) = 
\frac{\overline{P^2}}{2}\left(
\frac{2i\varGamma_0}{T(\omega+2i\varGamma_0)}
\left[1 - 2\frac{f^2R_\Sigma}{\pi\Theta}\int_{-\infty}^\infty dx\frac{|x|N(x)\Lambda(x)}{x+2i\varGamma_0}\right] 
+ 
2R_\Sigma\frac{f^2}{\pi\Theta}\int_{-\infty}^\infty dx \frac{{\rm sgn}(x)\Lambda(x)}{x+\omega+2i\varGamma_0}
\right),
\end{equation}
where we omit $\omega$ in the denominator of the integrand in the first term.

The quantity $\chi_2(\omega)$ can be expressed as follows:
\begin{eqnarray}
\label{chi22}
\chi_2(\omega) &=& J(\omega) + J^*(-\omega),\\
\label{j}
J(\omega) &=& \frac{\overline{P^2}}{2\pi} R_1\frac{f^2}{\pi\Theta} 
\int_{-\infty}^\infty dx dx_1 
e^{-x/T}|x_1|x_1N(x_1)\Lambda(x_1) G(x_1+x+\omega)G(x+\omega){\rm Im}\{G(x)G(x_1+x)\}.
\end{eqnarray}
The integral in Eq.~(\ref{j}) can be taken similar to those of $J_1(\omega)$ and $J_2(\omega)$, the result being
\begin{equation}
\label{chi2f}
\chi_2(\omega) = \frac{\overline{P^2}}{2}\left[
\frac{2i\varGamma_0}{T(\omega+2i\varGamma_0)}2R_1\frac{f^2}{\pi\Theta}\int_{-\infty}^\infty dx \frac{|x|N(x)\Lambda(x)}{x+2i\varGamma_0}
-
2R_1\frac{f^2}{\pi\Theta}\int_{-\infty}^\infty dx \frac{{\rm sgn}(x)\Lambda(x)}{x+\omega+2i\varGamma_0}
\right],
\end{equation}
where we omit $\omega$ in the denominator of the integrand in the first term.

The expression for $\chi_3(\omega)$ can be brought to the form:
\begin{equation}
\label{chi31}
\chi_3(\omega) = -\frac{\overline{P^2}}{2} \frac{\omega}{T(\omega+2i\varGamma)}Z_\Gamma
\int_{-\infty}^\infty dx
e^{-x/T}G(x+\omega)G^*(x)K(x).
\end{equation}
The area near poles of the Green's functions is essential in this integral. Therefore, we can replace $e^{-x/T}$ by unity and $K(x)$ by $K(0)=(\varGamma_0-\varGamma)/\pi$. As a result we have:
\begin{equation}
\label{chi32}
\chi_3(\omega) = \frac{\overline{P^2}}{2}
\left[
\frac{2i\varGamma}{T(\omega+2i\varGamma)}
-
\frac{2i\varGamma_0}{T(\omega+2i\varGamma_0)}
\right]
\left[
1
-
2R_\Sigma \frac{f^2}{\pi\Theta}\int_{-\infty}^\infty dx \frac{|x|N(x)\Lambda(x)}{x+2i\varGamma_0}
+
R_1 \frac{f^2}{\pi\Theta}\int_{-\infty}^\infty dx \frac{|x|N(x)\Lambda(x)}{x + 2i\varGamma_0}
\right].
\end{equation}
Summing Eqs.~(\ref{chi1}), (\ref{chi2f}) and (\ref{chi32}) we lead to Eq.~(\ref{chi2}) for the impurity susceptibility.

\section{Green's functions of 2D AF with impurities}
\label{gfwi}

We derive in this appendix Green's functions of operators $a$ and $a^\dagger$ considered in Appendix~\ref{delprop} in the case of 2D AF with impurities. To investigate the influence of the impurities we have to take into account the corresponding interaction (\ref{intaf}) in Dyson equations. It is assumed that $N,N_i\to\infty$ so as $N_i/N=n={\rm const}$, where $N$ and $N_i$ are the number of spins in the lattice and the number of impurities, respectively. Within the linear spin-wave approximation the equations for $g$ and $f^\dagger$ have the form
\begin{eqnarray}
\label{dysonnew}
g(\omega,{\bf k}) &=& g^{(0)}(\omega,{\bf k}) - g^2ns[1+\cos({\bf kR}_{12})]
[f^{(0)}(\omega,{\bf k}) + g^{(0)}(\omega,{\bf k})] \chi_x(\omega) [g(\omega,{\bf k}) + f^\dagger(\omega,{\bf k})] \nonumber\\
&&{} + g^2ns[1 - \cos({\bf kR}_{12})]
[f^{(0)}(\omega,{\bf k}) - g^{(0)}(\omega,{\bf k})] \chi_y(\omega) [g(\omega,{\bf k}) - f^\dagger(\omega,{\bf k})], \nonumber\\
f^\dagger(\omega,{\bf k}) &=& f^{\dagger(0)}(\omega,{\bf k}) - g^2ns[1+\cos({\bf kR}_{12})]
[\bar g^{(0)}(\omega,{\bf k}) + f^{\dagger(0)}(\omega,{\bf k})]\chi_x(\omega)[g(\omega,{\bf k}) + f^\dagger(\omega,{\bf k})]\nonumber\\
&&{} + g^2ns[1 - \cos({\bf kR}_{12})]
[\bar g^{(0)}(\omega,{\bf k}) - f^{\dagger(0)}(\omega,{\bf k})]\chi_y(\omega)[g(\omega,{\bf k}) - f^\dagger(\omega,{\bf k})],
\end{eqnarray}
where superscript $(0)$ denotes Green's functions without impurities which are given by Eq.~(\ref{gf}) and $\chi_x(\omega)$ and $\chi_y(\omega)$ are the impurity susceptibility given by Eq.~(\ref{chi3}) with $P=S^x$ and $S^y$, respectively. Equations (\ref{dysonnew}) can be easily solved with the result
\begin{eqnarray}
\label{gfi}
g(\omega,{\bf k}) &=& \bar g(-\omega,{\bf k})^* = \frac{sJ_0 + \omega - 2g^2ns\chi_\perp(\omega)}{{\cal D}(\omega,{\bf k})},
\nonumber\\
f^\dagger(\omega,{\bf k}) &=& f(-\omega,{\bf k})^* = \frac{-sJ_{\bf k} + 2g^2ns\chi_\perp(\omega)\cos({\bf kR}_{12})}{{\cal D}(\omega,{\bf k})},
\end{eqnarray}
where $\chi_\perp(\omega) = \chi_x(\omega) = \chi_y(\omega)$ and the denominator ${\cal D}(\omega,{\bf k})$ is given by Eq.~(\ref{sgfr}).

\bibliography{tls}

\begin{thebibliography}{39}
\expandafter\ifx\csname natexlab\endcsname\relax\def\natexlab#1{#1}\fi
\expandafter\ifx\csname bibnamefont\endcsname\relax
  \def\bibnamefont#1{#1}\fi
\expandafter\ifx\csname bibfnamefont\endcsname\relax
  \def\bibfnamefont#1{#1}\fi
\expandafter\ifx\csname citenamefont\endcsname\relax
  \def\citenamefont#1{#1}\fi
\expandafter\ifx\csname url\endcsname\relax
  \def\url#1{\texttt{#1}}\fi
\expandafter\ifx\csname urlprefix\endcsname\relax\def\urlprefix{URL }\fi
\providecommand{\bibinfo}[2]{#2}
\providecommand{\eprint}[2][]{\url{#2}}

\bibitem[{\citenamefont{Leggett et~al.}(1987)\citenamefont{Leggett,
  Chakravarty, Dorsey, Fisher, Garg, and Zwerger}}]{leggett}
\bibinfo{author}{\bibfnamefont{A.~J.} \bibnamefont{Leggett}},
  \bibinfo{author}{\bibfnamefont{S.}~\bibnamefont{Chakravarty}},
  \bibinfo{author}{\bibfnamefont{A.~T.} \bibnamefont{Dorsey}},
  \bibinfo{author}{\bibfnamefont{M.~P.~A.} \bibnamefont{Fisher}},
  \bibinfo{author}{\bibfnamefont{A.}~\bibnamefont{Garg}}, \bibnamefont{and}
  \bibinfo{author}{\bibfnamefont{W.}~\bibnamefont{Zwerger}},
  \bibinfo{journal}{Rev. Mod. Phys.} \textbf{\bibinfo{volume}{59}},
  \bibinfo{pages}{1} (\bibinfo{year}{1987}).

\bibitem[{\citenamefont{Weiss}(1999)}]{weiss}
\bibinfo{author}{\bibfnamefont{U.}~\bibnamefont{Weiss}},
  \emph{\bibinfo{title}{Quantum Dissipative Systems}}
  (\bibinfo{publisher}{World Scientific}, \bibinfo{address}{Singapore},
  \bibinfo{year}{1999}).

\bibitem[{\citenamefont{Maleyev}(1979)}]{mal2}
\bibinfo{author}{\bibfnamefont{S.~V.} \bibnamefont{Maleyev}},
  \bibinfo{journal}{Sov. Phys. JETP} \textbf{\bibinfo{volume}{52}},
  \bibinfo{pages}{1008} (\bibinfo{year}{1979}).

\bibitem[{\citenamefont{Abrikosov}(1965)}]{abrikos}
\bibinfo{author}{\bibfnamefont{A.~A.} \bibnamefont{Abrikosov}},
  \bibinfo{journal}{Physica} \textbf{\bibinfo{volume}{2}}, \bibinfo{pages}{5}
  (\bibinfo{year}{1965}).

\bibitem[{\citenamefont{Kokshenev}(1980)}]{kok}
\bibinfo{author}{\bibfnamefont{V.~B.} \bibnamefont{Kokshenev}},
  \bibinfo{journal}{Sov. Phys. JETP} \textbf{\bibinfo{volume}{6}},
  \bibinfo{pages}{1372} (\bibinfo{year}{1980}).

\bibitem[{\citenamefont{Maleyev}(1983)}]{mal4}
\bibinfo{author}{\bibfnamefont{S.~V.} \bibnamefont{Maleyev}},
  \bibinfo{journal}{Sov. Phys. JETP} \textbf{\bibinfo{volume}{57}},
  \bibinfo{pages}{149} (\bibinfo{year}{1983}).

\bibitem[{\citenamefont{Maleyev}(1994)}]{mal5}
\bibinfo{author}{\bibfnamefont{S.~V.} \bibnamefont{Maleyev}},
  \bibinfo{journal}{Phys. Rev. B} \textbf{\bibinfo{volume}{50}},
  \bibinfo{pages}{302} (\bibinfo{year}{1994}).

\bibitem[{\citenamefont{Mermin and Wagner}(1966)}]{mw}
\bibinfo{author}{\bibfnamefont{N.~D.} \bibnamefont{Mermin}} \bibnamefont{and}
  \bibinfo{author}{\bibfnamefont{F.}~\bibnamefont{Wagner}},
  \bibinfo{journal}{Phys. Rev. Lett.} \textbf{\bibinfo{volume}{17}},
  \bibinfo{pages}{1133} (\bibinfo{year}{1966}).

\bibitem[{\citenamefont{Kopietz}(1990)}]{kop}
\bibinfo{author}{\bibfnamefont{P.}~\bibnamefont{Kopietz}},
  \bibinfo{journal}{Phys. Rev. B} \textbf{\bibinfo{volume}{41}},
  \bibinfo{pages}{9228} (\bibinfo{year}{1990}).

\bibitem[{\citenamefont{Chakravarty et~al.}(1989)\citenamefont{Chakravarty,
  Halperin, and Nelson}}]{chak}
\bibinfo{author}{\bibfnamefont{S.}~\bibnamefont{Chakravarty}},
  \bibinfo{author}{\bibfnamefont{B.~I.} \bibnamefont{Halperin}},
  \bibnamefont{and} \bibinfo{author}{\bibfnamefont{D.~R.}
  \bibnamefont{Nelson}}, \bibinfo{journal}{Phys. Rev. B}
  \textbf{\bibinfo{volume}{39}}, \bibinfo{pages}{2344} (\bibinfo{year}{1989}).

\bibitem[{\citenamefont{Ty\v{c} and Halperin}(1990)}]{tyc}
\bibinfo{author}{\bibfnamefont{S.}~\bibnamefont{Ty\v{c}}} \bibnamefont{and}
  \bibinfo{author}{\bibfnamefont{B.~I.} \bibnamefont{Halperin}},
  \bibinfo{journal}{Phys. Rev. B} \textbf{\bibinfo{volume}{42}},
  \bibinfo{pages}{2096} (\bibinfo{year}{1990}).

\bibitem[{\citenamefont{Thurber et~al.}(1997)\citenamefont{Thurber, Hunt, Imai,
  Chou, and Lee}}]{thur}
\bibinfo{author}{\bibfnamefont{K.~R.} \bibnamefont{Thurber}},
  \bibinfo{author}{\bibfnamefont{A.~W.} \bibnamefont{Hunt}},
  \bibinfo{author}{\bibfnamefont{T.}~\bibnamefont{Imai}},
  \bibinfo{author}{\bibfnamefont{F.~C.} \bibnamefont{Chou}}, \bibnamefont{and}
  \bibinfo{author}{\bibfnamefont{Y.~S.} \bibnamefont{Lee}},
  \bibinfo{journal}{Phys. Rev. Lett.} \textbf{\bibinfo{volume}{79}},
  \bibinfo{pages}{171} (\bibinfo{year}{1997}).

\bibitem[{\citenamefont{Hoglund and Sandvik}(2003)}]{hog1}
\bibinfo{author}{\bibfnamefont{K.~H.} \bibnamefont{Hoglund}} \bibnamefont{and}
  \bibinfo{author}{\bibfnamefont{A.~W.} \bibnamefont{Sandvik}},
  \bibinfo{journal}{Phys. Rev. Lett.} \textbf{\bibinfo{volume}{91}},
  \bibinfo{pages}{077204} (\bibinfo{year}{2003}).

\bibitem[{\citenamefont{Hoglund and Sandvik}(2004)}]{hog2}
\bibinfo{author}{\bibfnamefont{K.~H.} \bibnamefont{Hoglund}} \bibnamefont{and}
  \bibinfo{author}{\bibfnamefont{A.~W.} \bibnamefont{Sandvik}},
  \bibinfo{journal}{Phys. Rev. B} \textbf{\bibinfo{volume}{70}},
  \bibinfo{pages}{024406} (\bibinfo{year}{2004}).

\bibitem[{\citenamefont{Sushkov}(2000)}]{sush1}
\bibinfo{author}{\bibfnamefont{O.~P.} \bibnamefont{Sushkov}},
  \bibinfo{journal}{Phys. Rev. B} \textbf{\bibinfo{volume}{62}},
  \bibinfo{pages}{12135} (\bibinfo{year}{2000}).

\bibitem[{\citenamefont{Sushkov}(2003)}]{sush2}
\bibinfo{author}{\bibfnamefont{O.~P.} \bibnamefont{Sushkov}},
  \bibinfo{journal}{Phys. Rev. B} \textbf{\bibinfo{volume}{68}},
  \bibinfo{pages}{094426} (\bibinfo{year}{2003}).

\bibitem[{\citenamefont{Nagaosa et~al.}(1989)\citenamefont{Nagaosa, Hatsugai,
  and Imada}}]{nagaosa}
\bibinfo{author}{\bibfnamefont{N.}~\bibnamefont{Nagaosa}},
  \bibinfo{author}{\bibfnamefont{Y.}~\bibnamefont{Hatsugai}}, \bibnamefont{and}
  \bibinfo{author}{\bibfnamefont{M.}~\bibnamefont{Imada}}, \bibinfo{journal}{J.
  Phys. Soc. Jpn.} \textbf{\bibinfo{volume}{58}}, \bibinfo{pages}{978}
  (\bibinfo{year}{1989}).

\bibitem[{\citenamefont{Igarashi et~al.}(1995)\citenamefont{Igarashi, Murayama,
  and Fulde}}]{igar}
\bibinfo{author}{\bibfnamefont{J.}~\bibnamefont{Igarashi}},
  \bibinfo{author}{\bibfnamefont{K.}~\bibnamefont{Murayama}}, \bibnamefont{and}
  \bibinfo{author}{\bibfnamefont{P.}~\bibnamefont{Fulde}},
  \bibinfo{journal}{Phys. Rev. B} \textbf{\bibinfo{volume}{52}},
  \bibinfo{pages}{15966} (\bibinfo{year}{1995}).

\bibitem[{\citenamefont{Murayama and Igarashi}(1996)}]{murayama}
\bibinfo{author}{\bibfnamefont{K.}~\bibnamefont{Murayama}} \bibnamefont{and}
  \bibinfo{author}{\bibfnamefont{J.}~\bibnamefont{Igarashi}},
  \bibinfo{journal}{J. Phys. Soc. Jpn.} \textbf{\bibinfo{volume}{66}},
  \bibinfo{pages}{1157} (\bibinfo{year}{1996}).

\bibitem[{\citenamefont{Clarke et~al.}(1993)\citenamefont{Clarke, Giamarchi,
  and Shraiman}}]{clarke}
\bibinfo{author}{\bibfnamefont{D.~G.} \bibnamefont{Clarke}},
  \bibinfo{author}{\bibfnamefont{T.}~\bibnamefont{Giamarchi}},
  \bibnamefont{and} \bibinfo{author}{\bibfnamefont{B.~I.}
  \bibnamefont{Shraiman}}, \bibinfo{journal}{Phys. Rev. B}
  \textbf{\bibinfo{volume}{48}}, \bibinfo{pages}{7070} (\bibinfo{year}{1993}).

\bibitem[{\citenamefont{Sachdev and Vojta}(2003)}]{sachdev}
\bibinfo{author}{\bibfnamefont{S.}~\bibnamefont{Sachdev}} \bibnamefont{and}
  \bibinfo{author}{\bibfnamefont{M.}~\bibnamefont{Vojta}},
  \bibinfo{journal}{Phys. Rev. B} \textbf{\bibinfo{volume}{68}},
  \bibinfo{pages}{064419} (\bibinfo{year}{2003}).

\bibitem[{\citenamefont{Vojta et~al.}(2000)\citenamefont{Vojta, Buragohain, and
  Sachdev}}]{vojta}
\bibinfo{author}{\bibfnamefont{M.}~\bibnamefont{Vojta}},
  \bibinfo{author}{\bibfnamefont{C.}~\bibnamefont{Buragohain}},
  \bibnamefont{and} \bibinfo{author}{\bibfnamefont{S.}~\bibnamefont{Sachdev}},
  \bibinfo{journal}{Phys. Rev. B} \textbf{\bibinfo{volume}{61}},
  \bibinfo{pages}{15152} (\bibinfo{year}{2000}).

\bibitem[{\citenamefont{Oitmaa et~al.}(1995)\citenamefont{Oitmaa, Betts, and
  Aydin}}]{oitmaa}
\bibinfo{author}{\bibfnamefont{J.}~\bibnamefont{Oitmaa}},
  \bibinfo{author}{\bibfnamefont{D.~D.} \bibnamefont{Betts}}, \bibnamefont{and}
  \bibinfo{author}{\bibfnamefont{M.}~\bibnamefont{Aydin}},
  \bibinfo{journal}{Phys. Rev. B} \textbf{\bibinfo{volume}{51}},
  \bibinfo{pages}{2896} (\bibinfo{year}{1995}).

\bibitem[{\citenamefont{Kotov et~al.}(1998)\citenamefont{Kotov, Oitmaa, and
  Sushkov}}]{kot2}
\bibinfo{author}{\bibfnamefont{V.~N.} \bibnamefont{Kotov}},
  \bibinfo{author}{\bibfnamefont{J.}~\bibnamefont{Oitmaa}}, \bibnamefont{and}
  \bibinfo{author}{\bibfnamefont{O.}~\bibnamefont{Sushkov}},
  \bibinfo{journal}{Phys. Rev. B} \textbf{\bibinfo{volume}{58}},
  \bibinfo{pages}{8500} (\bibinfo{year}{1998}).

\bibitem[{\citenamefont{Aharony et~al.}(1988)\citenamefont{Aharony, Birgeneau,
  Coniglio, Kastner, and Stanley}}]{aharony}
\bibinfo{author}{\bibfnamefont{A.}~\bibnamefont{Aharony}},
  \bibinfo{author}{\bibfnamefont{R.~J.} \bibnamefont{Birgeneau}},
  \bibinfo{author}{\bibfnamefont{A.}~\bibnamefont{Coniglio}},
  \bibinfo{author}{\bibfnamefont{M.~A.} \bibnamefont{Kastner}},
  \bibnamefont{and} \bibinfo{author}{\bibfnamefont{H.~E.}
  \bibnamefont{Stanley}}, \bibinfo{journal}{Phys. Rev. Lett.}
  \textbf{\bibinfo{volume}{60}}, \bibinfo{pages}{1330} (\bibinfo{year}{1988}).

\bibitem[{\citenamefont{Aristov and Maleyev}(1990)}]{aris}
\bibinfo{author}{\bibfnamefont{D.~N.} \bibnamefont{Aristov}} \bibnamefont{and}
  \bibinfo{author}{\bibfnamefont{S.~V.} \bibnamefont{Maleyev}},
  \bibinfo{journal}{Z. Phys. B} \textbf{\bibinfo{volume}{81}},
  \bibinfo{pages}{433} (\bibinfo{year}{1990}).

\bibitem[{\citenamefont{Chernyshov et~al.}(2002)\citenamefont{Chernyshov, Chen,
  and Neto}}]{chern}
\bibinfo{author}{\bibfnamefont{A.~L.} \bibnamefont{Chernyshov}},
  \bibinfo{author}{\bibfnamefont{Y.~C.} \bibnamefont{Chen}}, \bibnamefont{and}
  \bibinfo{author}{\bibfnamefont{A.~H.~C.} \bibnamefont{Neto}},
  \bibinfo{journal}{Phys. Rev. B} \textbf{\bibinfo{volume}{65}},
  \bibinfo{pages}{104407} (\bibinfo{year}{2002}).

\bibitem[{\citenamefont{Wan et~al.}(1993)\citenamefont{Wan, Harris, and
  Kumar}}]{wan}
\bibinfo{author}{\bibfnamefont{C.~C.} \bibnamefont{Wan}},
  \bibinfo{author}{\bibfnamefont{A.~B.} \bibnamefont{Harris}},
  \bibnamefont{and} \bibinfo{author}{\bibfnamefont{D.}~\bibnamefont{Kumar}},
  \bibinfo{journal}{Phys. Rev. B} \textbf{\bibinfo{volume}{48}},
  \bibinfo{pages}{1036} (\bibinfo{year}{1993}).

\bibitem[{\citenamefont{Mucciolo et~al.}(2004)\citenamefont{Mucciolo, Neto, and
  Chamon}}]{muc}
\bibinfo{author}{\bibfnamefont{E.~R.} \bibnamefont{Mucciolo}},
  \bibinfo{author}{\bibfnamefont{A.~H.~C.} \bibnamefont{Neto}},
  \bibnamefont{and} \bibinfo{author}{\bibfnamefont{C.}~\bibnamefont{Chamon}},
  \bibinfo{journal}{Phys. Rev. B} \textbf{\bibinfo{volume}{69}},
  \bibinfo{pages}{214424} (\bibinfo{year}{2004}).

\bibitem[{\citenamefont{Pirc and Dick}(1974)}]{pirc}
\bibinfo{author}{\bibfnamefont{R.}~\bibnamefont{Pirc}} \bibnamefont{and}
  \bibinfo{author}{\bibfnamefont{B.~G.} \bibnamefont{Dick}},
  \bibinfo{journal}{Phys. Rev. B} \textbf{\bibinfo{volume}{9}},
  \bibinfo{pages}{2701} (\bibinfo{year}{1974}).

\bibitem[{\citenamefont{Mao et~al.}(2003)\citenamefont{Mao, Coleman, Hooley,
  and Langreth}}]{major1}
\bibinfo{author}{\bibfnamefont{W.}~\bibnamefont{Mao}},
  \bibinfo{author}{\bibfnamefont{P.}~\bibnamefont{Coleman}},
  \bibinfo{author}{\bibfnamefont{C.}~\bibnamefont{Hooley}}, \bibnamefont{and}
  \bibinfo{author}{\bibfnamefont{D.}~\bibnamefont{Langreth}},
  \bibinfo{journal}{Phys. Rev. Lett.} \textbf{\bibinfo{volume}{91}},
  \bibinfo{pages}{207203} (\bibinfo{year}{2003}).

\bibitem[{\citenamefont{Shnirman and Makhlin}(2003)}]{major2}
\bibinfo{author}{\bibfnamefont{A.}~\bibnamefont{Shnirman}} \bibnamefont{and}
  \bibinfo{author}{\bibfnamefont{Y.}~\bibnamefont{Makhlin}},
  \bibinfo{journal}{Phys. Rev. Lett.} \textbf{\bibinfo{volume}{91}},
  \bibinfo{pages}{207204} (\bibinfo{year}{2003}).

\bibitem[{\citenamefont{Zawadovski and Fazekas}(1969)}]{zawadovski}
\bibinfo{author}{\bibfnamefont{A.}~\bibnamefont{Zawadovski}} \bibnamefont{and}
  \bibinfo{author}{\bibfnamefont{P.}~\bibnamefont{Fazekas}},
  \bibinfo{journal}{Z. Physik} \textbf{\bibinfo{volume}{226}},
  \bibinfo{pages}{235} (\bibinfo{year}{1969}).

\bibitem[{\citenamefont{Larsen}(1972)}]{larsen}
\bibinfo{author}{\bibfnamefont{U.}~\bibnamefont{Larsen}}, \bibinfo{journal}{Z.
  Physik} \textbf{\bibinfo{volume}{256}}, \bibinfo{pages}{65}
  (\bibinfo{year}{1972}).

\bibitem[{\citenamefont{Maleyev}(1970)}]{mal1}
\bibinfo{author}{\bibfnamefont{S.~V.} \bibnamefont{Maleyev}},
  \bibinfo{journal}{Teor. Mat. Fyz.} \textbf{\bibinfo{volume}{4}},
  \bibinfo{pages}{86} (\bibinfo{year}{1970}).

\bibitem[{\citenamefont{Ginzburg}(1974)}]{ginz}
\bibinfo{author}{\bibfnamefont{S.~L.} \bibnamefont{Ginzburg}},
  \bibinfo{journal}{Sov. Phys. Solid State} \textbf{\bibinfo{volume}{16}},
  \bibinfo{pages}{5} (\bibinfo{year}{1974}).

\bibitem[{\citenamefont{Maleyev}(2000)}]{malold}
\bibinfo{author}{\bibfnamefont{S.}~\bibnamefont{Maleyev}},
  \bibinfo{journal}{Phys. Rev. Lett.} \textbf{\bibinfo{volume}{85}},
  \bibinfo{pages}{3281} (\bibinfo{year}{2000}).

\bibitem[{\citenamefont{Petitgrand et~al.}(1999)\citenamefont{Petitgrand,
  Maleyev, Bourges, and Ivanov}}]{pet}
\bibinfo{author}{\bibfnamefont{D.}~\bibnamefont{Petitgrand}},
  \bibinfo{author}{\bibfnamefont{S.~V.} \bibnamefont{Maleyev}},
  \bibinfo{author}{\bibfnamefont{P.}~\bibnamefont{Bourges}}, \bibnamefont{and}
  \bibinfo{author}{\bibfnamefont{A.~S.} \bibnamefont{Ivanov}},
  \bibinfo{journal}{Phys. Rev. B} \textbf{\bibinfo{volume}{59}},
  \bibinfo{pages}{1079} (\bibinfo{year}{1999}).

\bibitem[{\citenamefont{Abrikosov et~al.}(1963)\citenamefont{Abrikosov,
  Gor'kov, and Dzyaloshinskii}}]{agd}
\bibinfo{author}{\bibfnamefont{A.~A.} \bibnamefont{Abrikosov}},
  \bibinfo{author}{\bibfnamefont{L.~P.} \bibnamefont{Gor'kov}},
  \bibnamefont{and} \bibinfo{author}{\bibfnamefont{I.~E.}
  \bibnamefont{Dzyaloshinskii}}, \emph{\bibinfo{title}{Quantum Field
  Theoretical Methods in Statistical Physics}} (\bibinfo{publisher}{Dover},
  \bibinfo{address}{New York}, \bibinfo{year}{1963}).

\end{thebibliography}

\newpage

\begin{figure}
\centering
\includegraphics[scale=0.9]{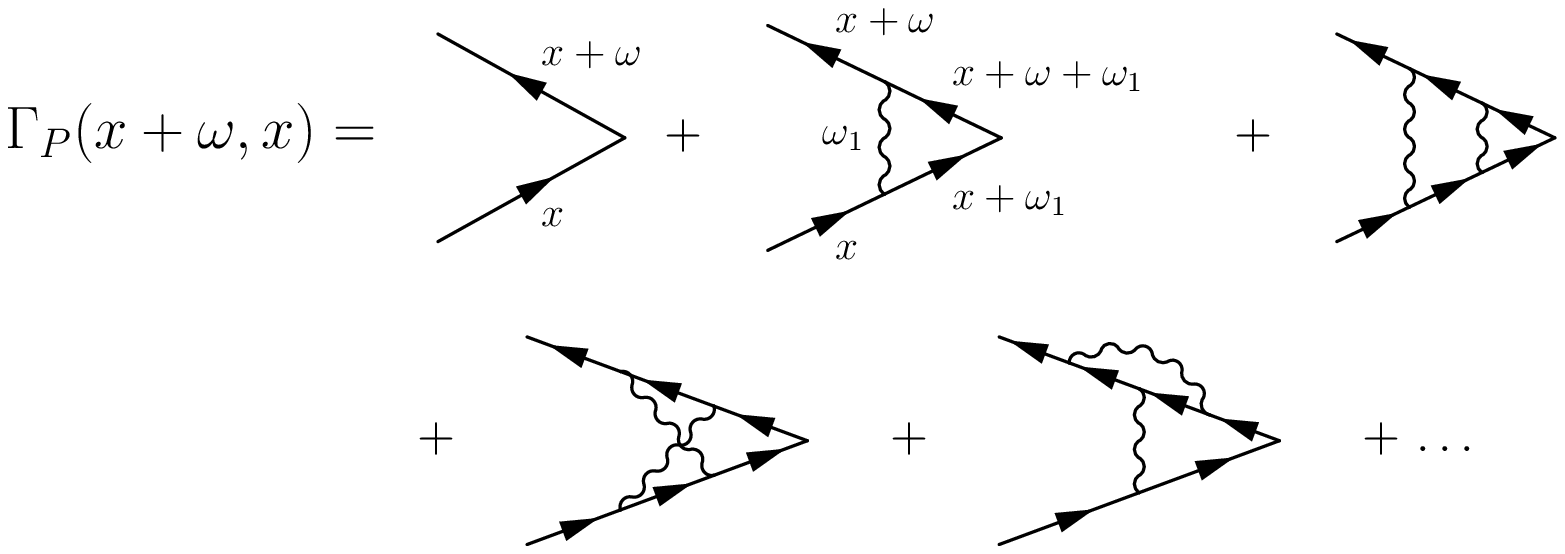}
\caption{The zero-, the first- and the second-order diagrams for the pseudofermion vertex $\Gamma_P (x+\omega,x)$.
\label{gammaf}} 
\end{figure}

\end{document}